\documentclass[a4paper,byrevtex,nofootinbib,twocolumn,showkeys,showpacs]{revtex4-1}
\usepackage[utf8]{inputenc}
\usepackage{siunitx}
\usepackage{booktabs}
\usepackage{graphicx}
\usepackage{amsmath}
\usepackage{amssymb}
\DeclareMathOperator{\sgn}{sign}

\newcommand{\us}{s}
\usepackage[autostyle]{csquotes}
\usepackage{url}
\usepackage{hyperref}

\usepackage{breakurl}
\graphicspath{{figure/}{../submissionSciRep/}}

\AtBeginDocument{
\heavyrulewidth=.08em
\lightrulewidth=.05em
\cmidrulewidth=.03em
\belowrulesep=.65ex
\belowbottomsep=0pt
\aboverulesep=.4ex
\abovetopsep=0pt
\cmidrulesep=\doublerulesep
\cmidrulekern=.5em
\defaultaddspace=.5em
}
\begin{document}

\title{Latitudinal trends in human primary activities: characterizing the winter day as a synchronizer}
\author{José María Martín-Olalla}
\affiliation{Universidad de Sevilla, Departamento de F{\'{\i}}sica de la Materia Condensada, PO Box 1065, ES41080 Seville, Spain}
\email{olalla@us.es; twitter: @MartinOlalla\_JM}

\begin{abstract}

This work analyzes time use surveys from 19 countries (17 European and 2 American) in the middle latitude range from \SIrange{38}{61}{\degree} latitude accounting for \SI{45}{\percent} of world population in that range. Time marks for primary activities (sleeping, working and eating) are systematically contrasted against light/dark conditions related to latitude. 

The analysis reveals that winter sunrise is a synchronizer for labor start time below $\SI{54}{\degree}$ where they occur within the winter civil twilight region. Winter sunset is a source of synchronization for labor end times.

Winter terminator also punctuate meal times in Europe with dinner times occurring \SI{3}{\hour} after winter sunset time within a strip of $\SI{1}{\hour}$, which is $\SI{40}{\percent}$ narrower than variability of dinner local times. 

The sleep-wake cycle of laborers in a weekday is shown to be related to winter sunrise whereas standard population's cycle appears to be irrespective of latitude.

The significance of the winter terminator depends on  two competing factors average daily labor time ($\sim$7h30m) and winter daytime ---the shortest photoperiod---. Winter terminator gains significance when shortest photoperiod roughly matches to daily labor time plus a reasonable lunch break. That is within a latitude range from \SIrange{38}{54}{\degree}.
 
The significance of winter terminator as a source of synchronization is also related to contemporary year round time schedules: the shortest photoperiod represents the worst case scenario the society faces.

Average daily sleep times show little trend with the shortest photoperiod slope $\sim\SI{5}{\min\per\hour}$ for a Pearson coefficient $r^2=0.242$. Average labor time may have a weak coupling with the shortest photoperiod: slope $\sim\SI{29}{\min\per\hour}$ for $r^2=0.338$.

\end{abstract}


\maketitle

\tableofcontents

\section{Introduction}
Earth's rotation period (one day or $T=\SI{24}{\hour}$, the definition of hour) notably influences biological rhythms\cite{Phillips2009,Kramer2013} and social human behavior. It is also the time basis to which mechanical clocks are sync. Clocks track time offset relative to the subsolar meridian ---the great circle intersecting Earth's rotation axis and the subsolar point, where the Sun is directly overhead--- hence clock time is insensitive to latitude. 

Earth's orbit and the obliquity of the rotation axis ($\epsilon=\ang{23.437}$) seasonally alters the latitude of the subsolar point and makes the terminator ---a great circle which separates light from darkness, with the subsolar point being one of its poles--- characteristically differ from meridians, except when an equinox occurs or when yearly averaged values are considered. As a consequence clock readings are out of sync with the terminator: along a meridian day and night depend on latitude. The terminator shapes sunrises sunsets, and generally speaking the seasonal light/dark cycle, which is best noticed in the middle latitude range: between tropics and polar regions.  

Human primary activity time marks (rise times, bed times, meal times, working times) are reported by the clock. It is an understanding that these time marks occur at prescribed values, within some variability linked to climate, cultural, legal, political or inherited habits. This understanding is a primary outcome of the worldwide use of time zones, which offset Earth's rotation and accommodate the variability solar events and human activities measured by an universal clock. An acceptable value of the variability of time marks linked to primary activities is one hour, the standard subdivision of one day and the standard width of a physical time zone. In so doing we inadvertently understand human activity is sync to noon and the terminator would play no role in the description of these time marks because it does not play a role in clock-time either.

The question to address in this work is to what extend is human primary activities sync to noon. Should people living along a meridian be doing these activities at the same time within some variability? An affirmative answer would result in meridional behavior of the activity. A negative answer would prompt a second question: would differences be systematic in latitude (the free parameter along a meridian)? Could they be linked to the non-meridional terminator? 

To some extend these questions aim to whether ancient time reckoning based on sunrises and sunsets and sensitive to the seasonal cycle still plays a role in industrialized societies highly tied to clocks synced to noon and whose only source of seasonality is Daylight Saving Time (DST). Sandford Fleming, as early as at the session held on Oct. 14, 1884 in the International Meridian Conference\cite{imc-1884}, put forward the idea that people can easily use clock time as a proxy for sunrise and sunset times by learning what the clock is ticking when they happen. At that moment Fleming was advocating for a universal clock worldwide but the idea still applies.

Daylight is the predominant synchronizer (zeitgeber) for the human clock.\cite{Roenneberg2013} Sleep activity can be traced by diurnal free preferences (circadian phenotypes or chronotype) which measure our preferred readiness by  morningness-eveningness tests like the Horne-Östberg (HO) questionnaire\cite{Horne1976}, the Composite Scale of Morningness\cite{Smith1989} (CSM) or the Munich ChronoType Questionnaire (MCTQ)\cite{Roenneberg2003}. For two of them (HO and CSM) the result is not given by a true metric of time, but a score. For MCTQ results are given through mid-sleep times, which are a true metric of time. Either case their scores and questionnaires refer to clock time, instead of time distance to the terminator line, which, understandably, is a complicated metric to set up in a questionnaire. Understanding the role of latitude\cite{White2003,Roenneberg2007,Randler2008,Miguel2014a} or latitude prone quantities like sunrise/sunset times, photoperiod or insolation, in morningness scores is then an open issue\cite{Leocadio-Miguel2017,Randler2017} with interesting derivatives.\cite{Monsivais2017}

This work is aimed to an analysis of human primary activities time marks extracted from time use surveys in seventeen European countries and two American countries which cover $\SI{45}{\percent}$ of the world population living in the middle latitude range from \SIrange{38}{61}{\degree}. Yearly averaged daily rhythms of main activities will be analyzed to obtain time marks representative of the country population. These time marks will be systematically contrasted against the light/dark daily and seasonal cycles. Not limited to free preferences or the sleep-wake cycle, the work is focused to labor activity which should be specifically prone to the light/dark cycle. The winter day ---that with the shortest photoperiod--- will show up as a synchronizer, within some variability, of human activity in this range of latitudes. The shortest photoperiod  ---the worst case scenario year round--- would force synchronization in modern societies with year round time schedules.

\section{Data sets}
\label{sec:dfa}

Time use surveys\cite{Chenu2006} are performed in many countries, chiefly OECD countries, with varying periodicity. Their aim is ascertaining when we do primary universal activities like sleeping, working or eating, and which fraction of a standard day an activity requires. Research on time budgets predates to late nineteenth century and has evolved so as to include comparisons among different classes of individuals\cite{craig2006}, the time evolution of societies\cite{aguiar2007} or its response to economic turmoils.\cite{aguiar2013}

Two sets of data will be studied in this paper. On the first hand, microdata from national time use surveys, six in Europe and two in America, which are publicly available\cite{dktus-2001e,estus-2010e,ustus-2012} or which could be obtained from institutions through petitions.\cite{frtus-2010e,ietus-2005,ittus-2010e,uktus-2003,catus-2005} European surveys were regionalized by NUTS\cite{eurostat-nuts} level 1 scheme; American surveys were regionalized by provinces (Canada) and Census Divisions (United States). However, this work will not analyze data from non-contiguous regions in France (overseas departments), Spain (Canary Islands) and United States. This paper will focus in primary activities (sleeping, working and eating) and the analysis will be enriched with the location ``at home'' and watching TV, an indoor leisure activity. 

In this set only laborers in a week day will be analyzed because their daily preferences should be most socially coupled and, also, should be driven most by external conditions like the light/dark cycle. Notice that this study is complementary to circadian phenotypes, which track free preferences. 

The number of respondents satisfying the condition lies in the range of five thousands within each participating country, except for Ireland (500), Denmark (1000) and United States, where a continuous, multiyear survey accounts for forty thousand respondents.  It will be assumed that the data provided by the survey represents yearly average conditions as it integrates answers throughout the year, except for Irish survey which did not expand along a calendar year. Respondents fill up a diary where a day is sliced into 144 time slots of ten minutes each, except Irish survey which make 96 time slots of one quarter of an hour each.

Highlighting differences between countries through time use surveys is a complicated issue because survey guidelines should be harmonized beforehand. That is the goal of the  Harmonised European Time Use Survey\cite{eurostat-hetus} (Hetus, here after) in Europe. The second set includes data retrieved from ``main activities/time of day'' pre-prepared tables available at Hetus\cite{hetus} webtool (\url{https://www.h6.scb.se/tus/tus/Statistics.html}) which accumulates results from surveys prior to 2005, many of them in the last decades of the 20th century. Here the standard population subset (twenty to seventy four years old) is analyzed. For the purpose of comparison with national time use surveys the shares of employees' in this subset ranges from \SIrange{34}{52}{\percent} with median at $\SI{43}{\percent}$. Sample size amounts to tens of thousands of respondents in most of the cases, ranging from 38000 (Italy) to 5500 (Norway). Also it should be mentioned that Hetus provide no data on locations; hence the location ``at home'' could not be retrieved in this set. 

In the following discussion geographical data of countries (latitude and longitude) will represent population weighted median values extracted from the database of cities with a population larger than 1000 inhabitants at \url{http://www.geonames.org}. The cast of countries geographically extends from \SIrange{38}{61}{\degree} in median latitude (see Table~\ref{tab:geo}). A Kolmogorov-Smirnov uniformity test does not reject ($p$-value=$\num{0.59}$) that the distribution of latitudes is uniformly distributed from minimum to maximum value, meaning voids in the distribution are unlikely. In 2010 roughly \SI{20}{\percent} of world population lived in that range on either hemisphere and the cast of countries to be analyzed accounts for \SI{45}{\percent} of them.\footnote{Population data from the Global Rural-Urban Mapping Project, compiled by William Rankin at Yale University \url{http://www.radicalcartography.net/index.html?histpop}}  The cast is culturally narrow: all of them are Western countries with Christian heritage in the Northern Hemisphere. With the sole exceptions of Lithuania and Bulgaria they belong to OECD. All European countries, except Norway, belong to the European Union. 
\section{Solar elevation angle and time}
\label{sec:geography}

Ambient light conditions are described by the the solar elevation angle $z$ ---the altitude of the Sun, the angle between the horizon and the center of the solar disc--- which is given by:\cite{pierre-cam-2014}
\begin{equation}
  \label{eq:1}
  \sin(z)=\cos(\lambda-\lambda_s)\cos\phi\cos\phi_s+\sin\phi\sin\phi_s,
\end{equation}
where $\phi,\lambda$  are the local latitude and  longitude, and $\phi_s,\lambda_s$ are subsolar latitude and longitude, where the Sun is overhead ($z=\ang{90}$) and no shadow is cast. Notice that in Equation~(\ref{eq:1}) the argument of the first cosine is the subsolar longitude offset relative to local longitude. 

Subsolar longitude changes with Earth's rotation at the rate of Earth's angular rotation of speed  $\Omega=2\pi/T=\SI{72.7}{\micro\radian\per\second}=\SI{15.0}{\degree\per\hour}$. The rotation speed scales longitude and time on Earth and allows to rewrite Equation~(\ref{eq:1}) as:
\begin{equation}
  \label{eq:4}
  \cos(\Omega\tau)=\frac{\sin(z)-\sin\phi\sin\phi_s}{\cos\phi\cos\phi_s},
\end{equation}
where $\tau$ is the time offset to local noon ($\lambda_s=\lambda, \tau=0$), usually known as mean solar time. 

Since the inverse cosine function is even, and provided that the absolute value of the RHS is smaller than one, Equation~(\ref{eq:4}) yields two solutions which are symmetrically disposed around $\tau=0$. One solution is the upward or morning solution; the other one is the downward or evening solution. The time elapsed between these solutions $D$ is then:
\begin{equation}
  \label{eq:9}
  D(z,\phi,\phi_s)=\frac{2}{\Omega}\cos^{-1}\left(\frac{\sin z-\sin\phi\sin\phi_s}{\cos\phi\cos\phi_s}\right).
\end{equation}
From this, upward and downward times can be expressed as:
\begin{equation}
  \label{eq:2}
  \tau_{\downarrow\uparrow}(z,\phi,\phi_s)=\frac{T\pm D(z,\phi,\phi_s)}{2},
\end{equation}
where $\tau$ is now conveniently set to $T/2$ (that is \SI{12}{\hour}) at noon.

For a critical elevation angle $z_c=\ang{-0.83}$, Equation~(\ref{eq:1}) gives the terminator; Equation~(\ref{eq:9}) gives daytime or photoperiod ---the time elapsed from solar upper limb crossing the horizon upward to crossing it back downward---, and Equation~(\ref{eq:2}) gives sunrise and sunset mean solar times. The critical $z_c$ is non-zero due to Sun's finite size and due to atmospheric refraction. The solution for $z_c=\ang{0}$ renders the terminator line as great circle and gently simplifies mathematics but will not be used here.

In the preceding equations the subsolar latitude changes yearly due to the obliquity of Earth's rotation axis, leads to the seasons of the year, and carries the calendar date in these equations. Solstices occur when subsolar latitude is the furthest to observer latitude in winter $\phi_s=-\sgn(\phi)\epsilon$ or the closest to observer latitude in summer $\phi_s=\sgn(\phi)\epsilon$ and, in this simple model, drive extreme solutions for the equations: the earliest and the latest sunrises and sunsets of the year, and the longest or shortest photoperiod. All of these extreme solutions strongly depend on local latitude at middle latitudes. In contrast, yearly averaged values, as well as  equinoctial ($\phi_s=0$) values, are nearly independent of the local latitude, except at polar latitudes.

Universal time (UT1) is conceptually a proxy for subsolar longitude to which is related by $T/2-\lambda_s/\Omega$, where $\lambda_s$ is given as an offset relative to UT1 prime meridian, the Greenwich meridian; UT1 is also mean solar time at this meridian. Local time $t$ differs from UT1 by the time zone offset $\Delta$ is usually given by a whole number of hours. Finally, mean solar time at any meridian differs from UT1 by $\lambda/\Omega$, which is the signed elapsed time between $\lambda_s=0$ and $\lambda_s=\lambda$ following Earth's rotation. As a result, mean solar time and local time are related by:

\begin{equation}
  \label{eq:6}
  \tau=t-\Delta+\frac{\lambda}{\Omega}=t-\delta,
\end{equation}
where the time offset $\delta$ measures how much local noon is delayed ($\delta>0$) or advanced ($\delta<0$) from civil midday. Time zone offset $\Delta$ is usually chosen so that $\delta$ lies in the range \SIrange{-30}{30}{\min} conforming a standard clock. However, in Belgium, France, Spain and the Canadian province of Saskatchewan (among others regions in the world) clocks are set one hour in advance so that $\delta$ ranges from \SIrange{30}{90}{\min}, a difference which must be taken into account when comparing time schedules and morningness-eveningness tests. Otherwise time schedules are reported abnormally delayed and morningness scores abnormally biased toward the eveningness. 

Table~\ref{tab:geo} lists the relevant geographical data for the participating countries and the forthcoming analysis: weighted population median latitude and time offset. It also lists latest sunrise time $t^\uparrow_w$ and earliest sunrise time $t^\downarrow_w$ year round expressed as local time. Finally, it lists the values of the shortest photoperiod $D_w(z_c,\phi,\sgn(\phi)\epsilon)$. Time values where computed using the equations in this section, fed with the listed values of latitude and time offset.

\section{Results}
\label{sec:results-1}

\subsection{Daily rhythms and relevant time marks}
\label{sec:activ-plots-relev}

\begin{figure*}
  \centering
  \includegraphics[width=\textwidth]{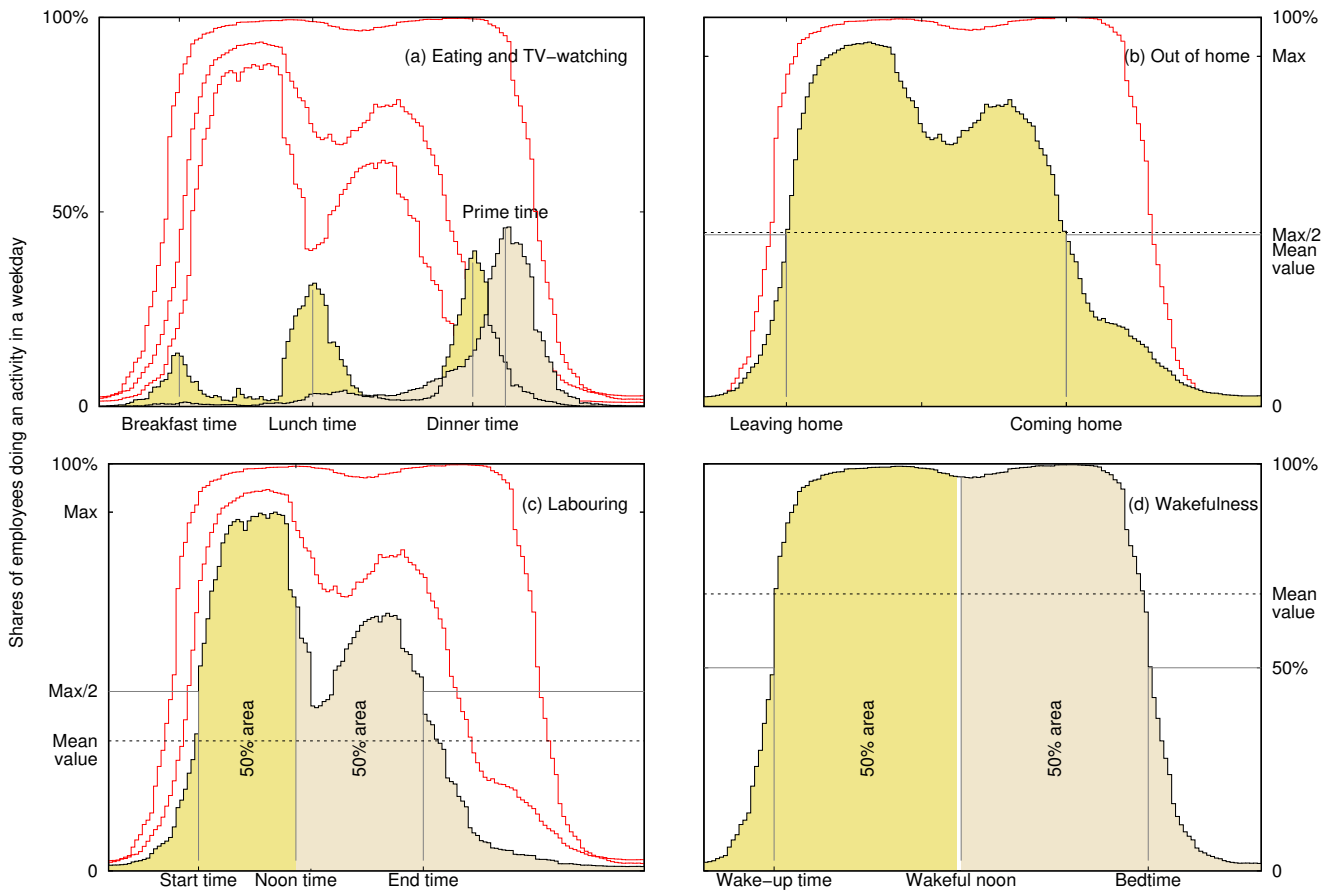}
  \caption{Daily rhythms and relevant time marks. Horizontal axes represent one day (\SI{24}{\hour}) starting at 4am local time and vertical axes  the shares of employees doing an activity at a given time on a weekday (Monday to Friday) in Italy as an example. From bottom to top and right to left: panel (d) shows the sleep-wake daily rhythm in filled style. Wakeful noon is the instant when half  area has been consumed and half remains. Panel (b) mimics panel (d) but refers to the location ``out of home''. The threshold defining leaving and coming home times are set to one half of the maximum value of the daily rhythm.  The sleep-wake daily rhythm is displayed unfilled to provide context. Panel (c) shows in filled style the labor daily rhythm with the sleep-wake and out of home daily rhythms shown unfilled to provide context. Panel (a) shows eating and TV-watching daily rhythm in filled style; the remaining daily rhythms in unfilled style provide context. In panel (a) relevant time marks are determined from peak positions.}

 \label{fig:activida}
\end{figure*}

Relevant time marks for the budget of time can be retrieved from the so-called daily rhythms where the shares of the sample (or a specific subset of the sample) doing a prescribed activity are shown as a function of time on an average day. The daily rhythm provide an insight into activity rate performance. It integrates regionally, hourly, daily and yearly the myriad of decisions that shape the lives of individuals. It is an implicit understanding that these decisions are independent between national time use surveys and it is this country level average values that will be compared in the forthcoming analysis. Regional values, tied by social clock within a country, will not be compared.

As an example Figure~\ref{fig:activida} shows in fill style daily rhythms obtained from the Italian time use survey starting at 4am and ending \SI{24}{\hour} later. The shaded area is the average daily consumption of the activity: the plot area amounts to one whole day in each panel. The mean value (panels b, c and d) is the shares of employees that would have lead to the equivalent consumption if the activity were performed steadily, which is never the case. Mean value also measures daily consumption as a fraction of one day. The change of a daily rhythm with time shows net flux of people starting and stopping doing an activity at that time.

Start and end times can be computed on panels (b), (c) and (d) with the help of a threshold located at half the daily maximum rhythm. They characterize when a society is activated or deactivated from the point of view of an activity. On panel (a), eating and TV-watching activities occur in bursts. Peak positions will identify the relevant times. Eating have some specificity's: first, surveys can not track energy intake so there is no difference in data between taking a snack or a proper meal. Secondly, in contrast with the daily rhythm shown on panel (a), the shares of population eating at a peak can be low enough so that the rhythm shows saw-tooth behavior, showing the preference for discrete, whole number of hours. For that reason eating daily rhythms coming from the Hetus webtool have all been smoothed and filtered with a Butterworth low-pass filter. Peaks were determined for the filtered signal. A visual inspection suffices to test that the algorithm retrieve well a time representative of each meal for every country.

A third time-mark will characterize labor and sleep-wake activity (panels c and d): activity noon time will be defined as the instant of time when the half daily consumption has been burned and half remains. Wakeful noon is twelve hour apart from sleep noon, which slightly differ from mid-sleep.

The sleep-wake daily rhythm (panel d) resembles a rectangular function with two states (wakeful and asleep). Albeit for the statistical rounding effects at wake-up time and bedtime, sleep-wake daily rhythm is not different from the wakeful state of many individuals. This shows a confluence of individual decisions which is a primary observation of daylight (also a rectangular function) as a zeitgeber. 

A fraction of awaken employees are ``out of home''\footnote{Some surveys do not release location data for the sleeping activity to protect privacy. In those cases the sleeping activity has been assumed to be located at home as default. Therefore, strictly speaking these data read being at home or sleeping.} (panel b) at any hour of the day. Panel (b) shows this daily rhythm is sometimes far from a rectangular function: laborers can be at or out of home in a range of times and due to a myriad of reasons. One of these reasons is working (see panel c) whose daily rhythm is even more complex as it integrates morning, afternoon, night and split shifts in different ways. 

In the morning, and quite generally, sleep-wake, out of home and labor daily rhythm soars very fast and with little time distance between them: employees get up, have breakfast, leave home and rush to get to work in a little time distance (see panel (c)). However, in the afternoon, the reverse process characteristically widens in time: each daily rhythm decreases at slower pace and their plunges are separated in time. That is an interesting difference with solar activity, which is perfectly symmetric around noon (see Equation~(\ref{eq:1})) on a daily basis.



\subsection{Time marks and latitude}
\label{sec:latitudinal-trends}

A comprehensive analysis of human activity in relation to the returning phenomena of light and darkness is an open issue which must be related to longitude and latitude. Whereas the role of longitude is understood by converting local times into solar times through Equation~(\ref{eq:6}), the role of latitude is more complex.

\begin{figure*}
  \centering
  \includegraphics[width=\textwidth]{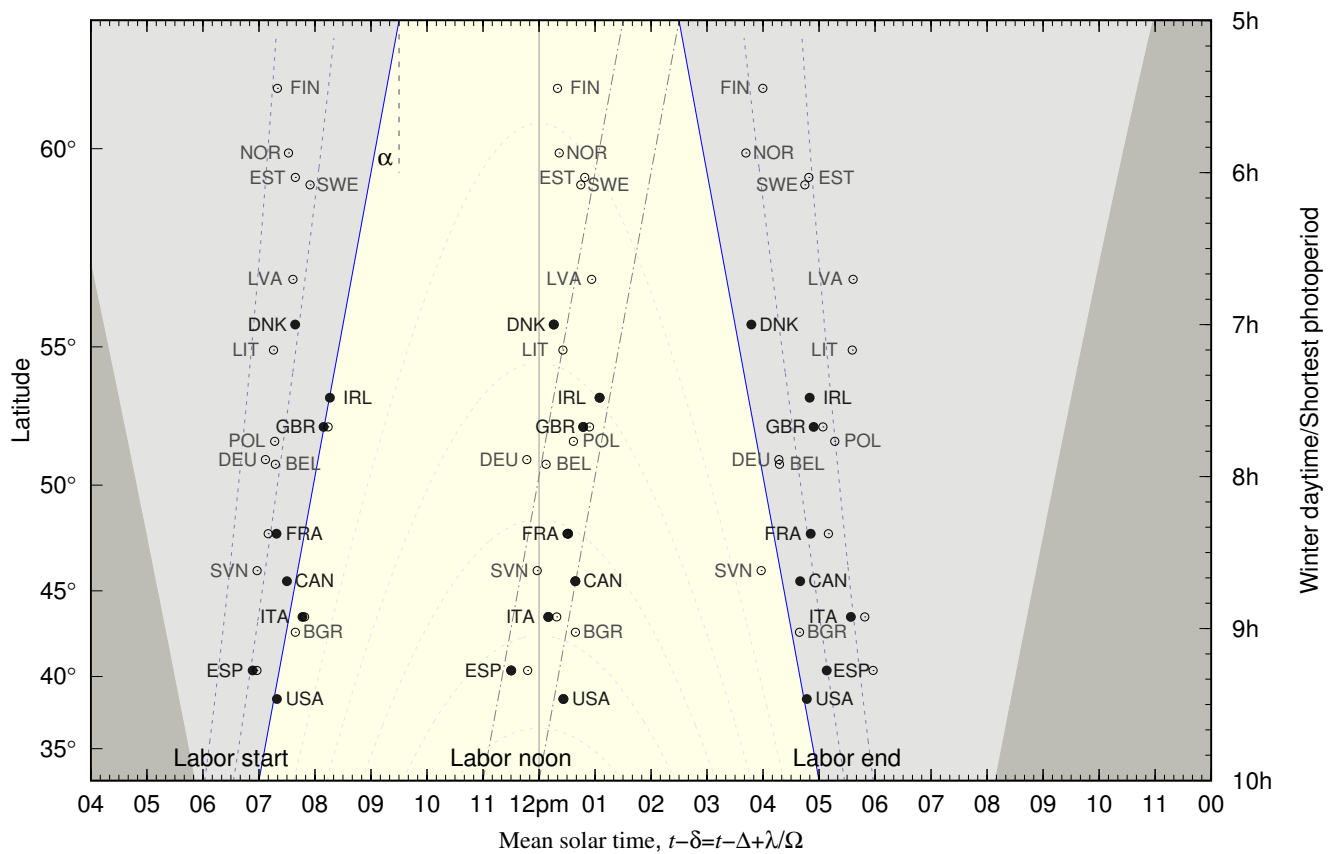}

  \caption{The daily and seasonal light/dark cycle and labor time marks versus winter daytime or latitude. Solid symbol stands for data extracted from national time use surveys; open symbols, for data computed from Hetus. Background colors display ambient light conditions. The lightest background displays the region where $z>z_c$ (the solar limb is above local horizon) irrespective of calendar day. The darkest background displays the region where darkness prevails irrespective of calendar day (artificially delayed by one hour to accommodate daylight saving time). In the intermediate background color light and darkness depends on calendar date. Dashed lines shows a grid of constant $z$ lines valid for the winter solstice and separated by $\ang{6}$ starting at $\ang{-12}$ (nautical twilight) which is the outer most in the morning and evening. The slope $\alpha=-1/2=-\SI{30}{\min\per\hour}$ for the winter sunrise time vs shortest photoperiod is noted. Dash-dotted straight lines shows \SIrange{4}{5}{\hour} after winter sunrise. Labels indicate ISO-3166-1 alpha-3 country codes. Data are listed in Table~\ref{tab:data}. Bivariate correlations are reported in Table~\ref{tab:ajustelabor}.}
  \label{fig:labor}
\end{figure*}

Figure~\ref{fig:labor} shows in graphic style the solutions to Equations~(\ref{eq:4}) and~(\ref{eq:2}) which shape ambient light conditions daily and year round. Notice that $x$-axis, mean solar time, accounts for longitude and Earth's rotation collapsing both in a common reference where noon happens 12pm (noted by a vertical line) irrespective of latitude. Vertical axes display latitude (left) and straight values of winter daytime (right) $D_w=D(z_c,\phi,-\sgn(\phi)\epsilon)$ (see Equation~(\ref{eq:9})) which is the shortest photoperiod year round for a given latitudinal cline. In the forthcoming analysis $D_w$ will play the role of a proxy for latitude. Notice that $D_w$ decreases on increasing $\phi$ so that the right axis runs upside-down from the highest to the lowest value.

Background colors display ambient light conditions. The lightest background shows the region where light prevails irrespective of calendar date. It is bounded by the winter terminator, which is piece-wise linear in winter daytime (see Equation~(\ref{eq:2})) with slopes $\alpha=\SI{\pm30}{\min\per\hour}$ (minutes of change in sunset or sunrise time per hour of change in photoperiod), a value to keep in mind. Seemingly, darkest background shows the region bounded by the summer terminator\footnote{The summer terminator has been delayed by one hour to simulate daylight saving time (a biannual change in $\Delta$), which is enforced in the regions analyzed in this paper with exceptions of Saskatchewan (Canada) and Arizona (United States).} which sets the shortest night year round. Intermediate background displays region where light or darkness seasonally alternates. Straight values of $D_w$ make these regions have polygonal shape. Straight values of $\phi$ would have rendered them curvilinear as in any web page that shows a day/night map of the Earth.

Dashed lines display constant solar elevation angle at winter $z_w$. Following Equation~(\ref{eq:4}) noon and midnight are axes of symmetry for them. The outer most line is $z_w=\ang{-12}$ ---where nautical twilight starts or ends--- followed by $z_w=\ang{-6}$ ---where winter civil twilight starts or ends. They  characterize the significant change in ambient light conditions at dawn and dusk. The solid line in blue shows the terminator $z_w=z_c=\ang{-0.83}$. Afterwards lines are spaced by $\Delta z_w=\ang{6}$ starting at $z_w=\ang{6}$.


Upon this natural framework which shows up the daily and seasonal cycle of light and darkness Figure~\ref{fig:labor} display labor time marks whose values are listed on Table~\ref{tab:data}.

While data show some apparent trends latitude, it is placement the most significant feature of labor start and end time marks. They are placed in the neighborhood of the winter terminator despite they were obtained from yearly averaged labor daily rhythms. Notice that winter sunrise and summer sunset are just $\SI{12}{\hour}$ (solar time) apart, the same applies to summer sunrise and winter sunset. Therefore it is placement, and the significance of either photoperiod, that allows to discriminate winter sunrise trend from summer sunset trend.

 Below $\SI{54}{\degree}$ latitude (see Table~\ref{tab:geo}) a bunch of labor start times are located around the winter civil twilight line and another pocket are located at the winter terminator. As a whole most of them are boxed within the winter civil twilight region $z\in(-\ang{6},\ang{0}$). Above $\SI{54}{\degree}$ start labor times are boxed in the winter nautical region $z\in(-\ang{12},-\ang{6})$. End times are boxed in a wider region of $\sim\ang{12}$. Table~\ref{tab:data} lists $z_w$ values for labor start and end times. The significance of these placements lie in the fact that they are describing the weakest values for $z$ at labor start and end times year round.

Labor noon times are located close to solar noon yet a trend with latitude is apparent and data below $\ang{54}$ latitude can be boxed on one-hour strip centered 4h30m after winter sunrise. 

Placement and boxing will be a key technique for understanding data but they do not exclude correlations For this two strategies will be followed. First, bivariate correlations $\tau$ vs $D_w$ will be tested. Non-zero Pearson's $r^2$ and slopes $p$ will punctuate a trend with $D_w$, meaning with latitude. Also variability ---measured as twice the sample standard deviation $2\us(\{\tau_i\})$--- will be of interest, although this is not a property of the trend but a descriptive parameter of the distribution $\{\tau_i\}$. 

The second strategy will test bivariate correlations for a latitudinal property ---for instance $z_w$--- at time marks vs $D_w$. Small variability followed with lower values of Pearson's coefficient and slope will show the tested property is stationary with latitude.

Table~\ref{tab:ajustelabor} reports bivariate correlations for labor time marks. Testing against $\tau$ results in non-zero but low Pearson's $r^2=\{0.233,0.277,0.0845\}$ (start,noon,end) resulting in large slope uncertainties. Weak trends of $\tau$ with latitude are indicating that winter terminator is not the only source of the variance in labor times. Cultural issues still play a role including preferences for whole hours. 

Slopes show increasing value of start and noon times with decreasing values of $D_w$ and increasing values of end times with increasing values of $D_w$. For labor start noon times $p\sim-\SI{20}{\min\per\hour}$ which is $2/3$ of $\alpha$ (the slope of a one-to-one correspondence of labor start times and winter sunrise time). That means data lean toward earlier times as latitude increases, like twilight lines do.

The latitude prone quantity to be tested will be $z_w$ for labor start and end times and distance to winter sunrise $\Delta t_w$ for labor noon. Pearson's $r^2$ are largely reduced $r^2=\{0.045,0.053,0.021\}$. Uncertainties cover the zero-slope case. Labor noon time variability decreases  $\SI{20}{\percent}$ ($\SI{10}{\min}$).  The variability of $z_w$ is one twilight region (start) and one and a half (end).

All these evidences characterize the behavior of labor start and end time marks with latitude in the range $\ang{35}$ to $\ang{54}$. The winter terminator arises as a synchronizer for labor start and end times in this range. Start times seems to be stronger synchronized to winter sunrise than end times do to winter sunset. However it should be mentioned a major problem this analysis faces: the span of the independent variable $D_w$  (1h50m, see Table~\ref{tab:geo}) is similar to the default variability in the dependent variable $\tau$.

Above $\SI{54}{\degree}$ the scenario differs: twilight zone opens and start becoming indistinguishable from meridians. Also $D_w$ falls well below the average labor time and sample size is small. Table~\ref{tab:ajustelabor} reports bivariate correlations for labor times above $\ang{54}$. They apparently behave meridionally. Yet placement is remarkably located in the winter nautical region (start times).

\begin{figure*}
  \centering
  \includegraphics[width=\textwidth]{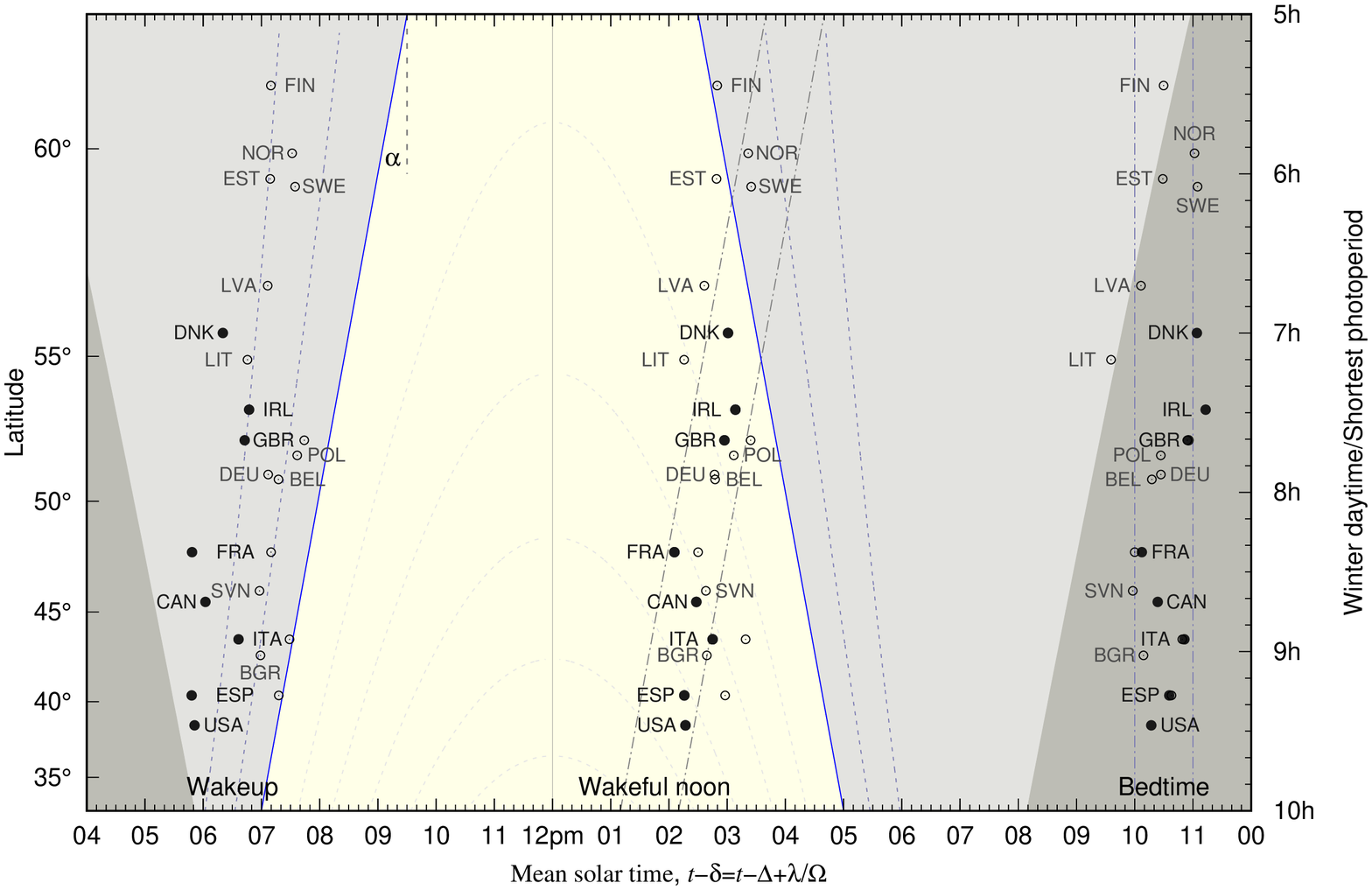}
  \includegraphics[width=\textwidth]{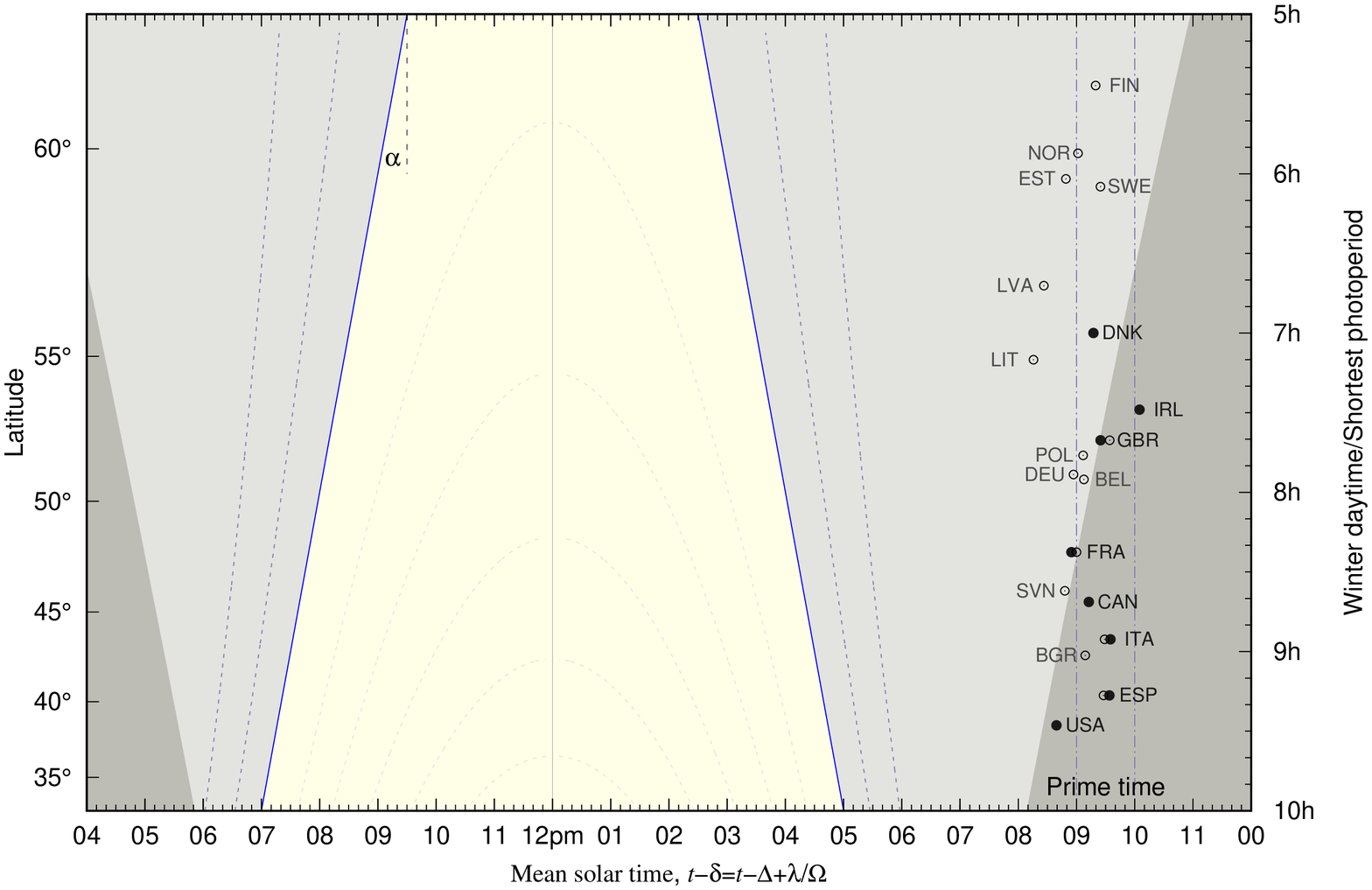}
  \caption{The same background as in Figure~\ref{fig:labor} and the sleep-wake solar time marks versus latitude. From left to right: wake-up time, wakeful noon and bedtime. Solid symbols show data extracted from time use surveys for employees only; open symbols data from Hetus for the standard population. The slanted dash-dotted lines display a strip one hour centered 6h35m after winter sunrise. Data are listed in Table~\ref{tab:vigil}. Bivariate correlations are reported in Table~\ref{tab:ajusteVigil}. Bottom figure shows TV prime time markes, data listed in Table~\ref{tab:ociodata} and bivariate correlations reported in Table~\ref{tab:ajusteTV}.}
  \label{fig:vigil}
\end{figure*}

These analyses can be mimicked for the remaining primary activities. Table~\ref{tab:vigil} lists time marks for the sleep-wake cycle. Data are shown in figure~\ref{fig:vigil}. Bivariate correlation data are reported in Table~\ref{tab:ajusteVigil}.

Some complexities must be noted for this activity. First bold symbols refer to employees' statistics while light symbols refer to standard population statistics. The standard population has a varying shares of non-employees within each country ranging from \SIrange{48}{66}{\percent} of sample size. Bold and light symbols largely differ in wake up times but not on bedtimes. Comparison of French, Italian, British and Spanish data, which appear on both sets, is relevant. 

It is also an issue that Hetus webtool does not provide the sleep-wake daily rhythm. Instead it provides the ``sleep and other personal care'' daily rhythm. On this circumstance one should expect that start time marks would be slightly delayed: if an individual wake up and start taking personal care ---for instance, washing up--- she would not have changed her status in the daily rhythm until she finished doing personal care. For the same reason end times may be slightly advanced if individuals perform personal care just before bedtime. 

Bivariate correlations $\tau$ vs $D_w$ for employees' wake up times report $r^2=\num{0.407}$ and $p=\SI{-17.8(88)}{\min\per\hour}$, roughly $3/5$ of $\alpha$. Correlations $z_w$ vs $D_w$ show smaller correlation $r^2=\num{0.0292}$ with $z_w$ irrespective of latitude and close to $\SI{-15}{\degree}$ and variability equation to one twilight region ($\ang{6}$). Wake up conditions for employees are largely related to labor start times, synchronized by the winter sunrise. Values of $z_w$ at wake up times are listed in Table~\ref{tab:vigil}.

Bivariate correlations $\tau$ vs $D_w$ for wakeful noon and bedtimes give similar $r^2$ and $p$ to those of wake up times suggesting that the morning trend is kept throughout the sleep-wake cycle. Wakeful noon times can be boxed in an one hour strip centered 6h40m after sunrise (shown in Figure~\ref{fig:vigil}); bivariate correlations $\Delta t_w$ vs $D_w$ shrinks $r^2$ and $2\us(\{y_i\})$. Same results are obtained for bedtimes, located 9h10m before winter sunrise.

Bivariate correlations for standard population (see Table~\ref{tab:ajusteVigil}) tell a different history: wake up, wakeful noon and bedtimes show little correlation in $\tau$ vs $D_w$ analysis with $r^2=\{0.013,0.028,0.07\}$ and slopes $1/6$ to $1/30$ smaller than $\alpha$ and negative: time marks get slightly delayed as latitude increases. This is suggesting that sleeping time in standard population are largely irrespective of latitude. 

Notwithstanding this, visual inspections of bedtimes in Table~\ref{tab:vigil} or Figure~\ref{fig:vigil} suggest that they are similar for both sets (employees and standard population) and both set of data can be boxed in a one-hour strip located at 10pm-11pm advocating that they may be globally irrespective of latitude. Bivariate correlations $\tau$ vs $D_w$ for the full set of data result in low Pearson's coefficient $r^2=\num{0.0466}$. In any case the hypotheses concerning employees' bedtimes face the problem of low sample size in this analysis.

Table~\ref{tab:ociodata} display data for TV prime time marks which occur in the hour preceding bedtimes. Time marks are similar for laborers and standard population. Table~\ref{tab:ajusteTV} reports data on bivariate correlations $D_w$ vs $\tau$ and the full set of observations which display $r^2=\num{0.0342}$ and small slope compared to $\alpha$. TV prime time marks are then irrespective of latitude, a result which also applies for employees' only data. This indirectly supports the hypothesis of bedtimes irrespective of latitude.

Relevant times for the location ``out of home'' and employees are listed in Table~\ref{tab:casa}. Leaving home time mark occurs at the end of nautical winter twilight and is still correlated to labor start time: laborers readily get up, leave home and get to work in the early hours of the morning. 

Coming home time marks spans along some hours when expressed in local time (see Table~\ref{tab:casa}). Contrastingly, they lie within a one-hour strip centered at 2h10m after winter sunset; European data only span through $\SI{45}{\minute}$. Table~\ref{tab:ajusteCasa} reports the statistics for bivariate correlations between the location ``out of home'' with distance to winter sunset presenting best results for this small data set. Slope is negligible and variability in $\Delta t_w$ is $\SI{40}{\percent}$ ($\SI{30}{\min}$) smaller than variability in $\tau$. Distance to winter sunset values are listed in Table~\ref{tab:casa}. Also with leaving home times linked to labor start the winter terminator seems to synchronize this activity. 

\begin{figure*}
  \centering
  \includegraphics[width=\textwidth]{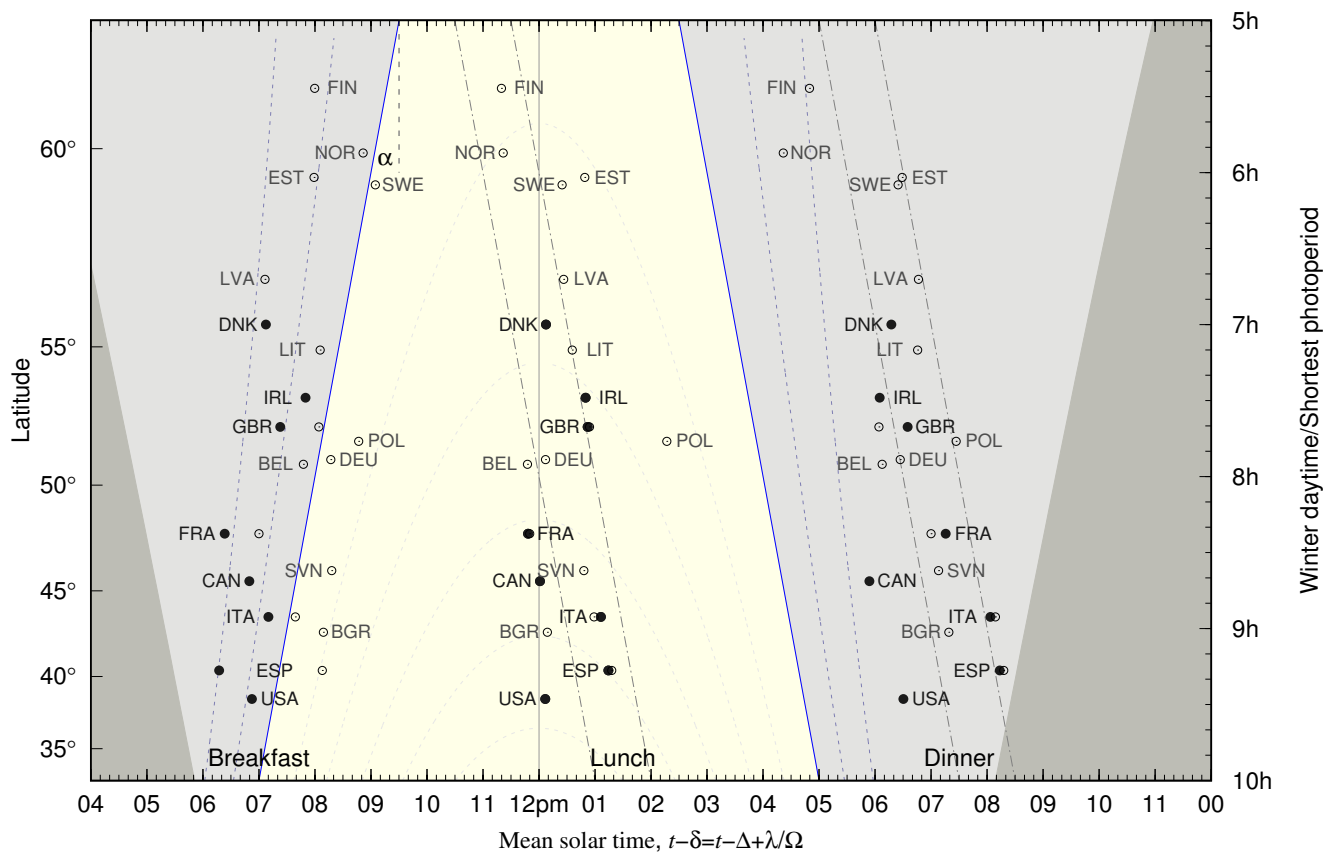}

  \caption{The same as in Figure~\ref{fig:labor} or~\ref{fig:vigil} but with eating time marks: breakfast (left), lunch (center) and dinner (right). Solid symbols show data extracted from time use surveys and refer to laborers in a week day. Open symbols (Hetus) refer to the standard set of population. Dash-dotted lines display two bands of one hour width. The earliest band is centered at three and half hours before winter sunset. The latest band is centered at three hours after winter sunset. They enlighten the comparison of $\Delta t_{\min}$ in Table~\ref{tab:meal}. Bivariate correlations are reported on Table~\ref{tab:ajusteComida}.}
  \label{fig:meal}
\end{figure*}

Finally, Figure~\ref{fig:meal} shows the relevant times associated with the eating daily rhythm. Breakfast times are largely different for solid symbols (employees) and open symbols (standard population) in the few cases for which both statistics have been analyzed. In contrast, time marks for lunch and dinner apparently do not differentiate one from the other. Both of them exhibits an interesting pattern with latitude.

 Dinner times occur well in the winter night from 6pm to 8pm mean solar time. European dinning times, save for Norwegian time, lie in the neighborhood of a one-hour strip centered at $\SI{3}{\hour}$ after winter sunset, shown in figure~\ref{fig:meal}. In contrast American dinner times tend to occur only two hours after winter sunset. Table~\ref{tab:meal} lists values of distance to winter terminator for meal times.

Table~\ref{tab:ajusteComida} reports the bivariate correlations for European data only, excluding Norway. Testing $\tau$ against $D_w$ result in large correlations ($r^2=\num{0.659}$) with slope $\SI{38.4(65)}{\min\per\hour}$ ($4/15$ excess of $\alpha$) and accounting for the variability in dinning times $2\us\sim\SI{105}{\min}$, which get delayed as latitude decreases. Pearson's coefficient for dinning time is gently reduced when testing against $\Delta t_w$: $r^2=\num{0.0859}$. Data collapse in the one-hour strip shown in figure~\ref{fig:meal} and variability is reduced by $\SI{40}{\percent}$ to $2\us\sim\SI{65}{\min}$. All of this suggests winter sunset could be a source of synchronization for dinning times.


Lunch times, on their hand, can be understood upon two premises. From one point of view lunch times can be related to solar noon, irrespective of latitude. Some of the data point well stick to the 12pm vertical line, interestingly American data (see Figure~\ref{fig:meal}). Those that do not stick to this line, stick to the 1pm vertical line, which simulates noon time in summer, due to daylight saving time. In this scenario lunch times would be a meridional.

Yet, from another point of view, many European data are placed into a one-hour strip centered 3h30m before winter sunset. In this second scenario lunch time would come to mean eating well before sunset or with still enough ambient light conditions. 

Interestingly noon intersects this strip from $\SI{50}{\degree}$ to $\SI{60}{\degree}$ so that having lunch at noon or three hours before sunset means roughly the same at that range. At $\SI{60}{\degree}$ latitude people may anticipate lunch as a way of foreseeing the winter sunset. On the contrary at $\SI{40}{\degree}$ people can delay lunch times since winter sunset is also delayed.

Table~\ref{tab:ajusteComida} reports similar numbers for bivariate correlations $\tau$ vs $D_w$ and $\Delta t_w$ vs $D_w$ at lunch time and European data, excluding Poland. This is likely indicating that both synchronizers ---noon and winter sunset--- are equally observed in the set of observations.  

\subsection{Average daily consumptions}
\label{sec:latit-trends-daily}

Average daily consumption falls in another category of data. They are not related to clock-time, time zones and longitude, but to stopwatch: a duration of time. They represent the shares of one day consumed by an activity.

The shaded area in Fig.~\ref{fig:activida} is the average daily consumption of every activity: it integrates every daily contribution to the activity. For the sleep-wake cycle notice that sleep time is not the right straight difference of wake up times and bedtimes: it also includes intra-day sleeping activity (siesta time) and the different paces at which sleep-wake cycle soars in the morning and drops at night. Table~\ref{tab:duracion} lists the daily average consumption of the primary activities labor, eating and sleeping ---the complementary of the wakeful consumption shown in panel (d) of figure~\ref{fig:activida}.

Bivariate correlations for daily consumptions and $D_w$ are reported on Table~\ref{tab:ajusteDuracion} for each data subset and both combined; correlations on daily TV time are also reported. 

\begin{figure*}
  \centering
  \includegraphics[width=\textwidth]{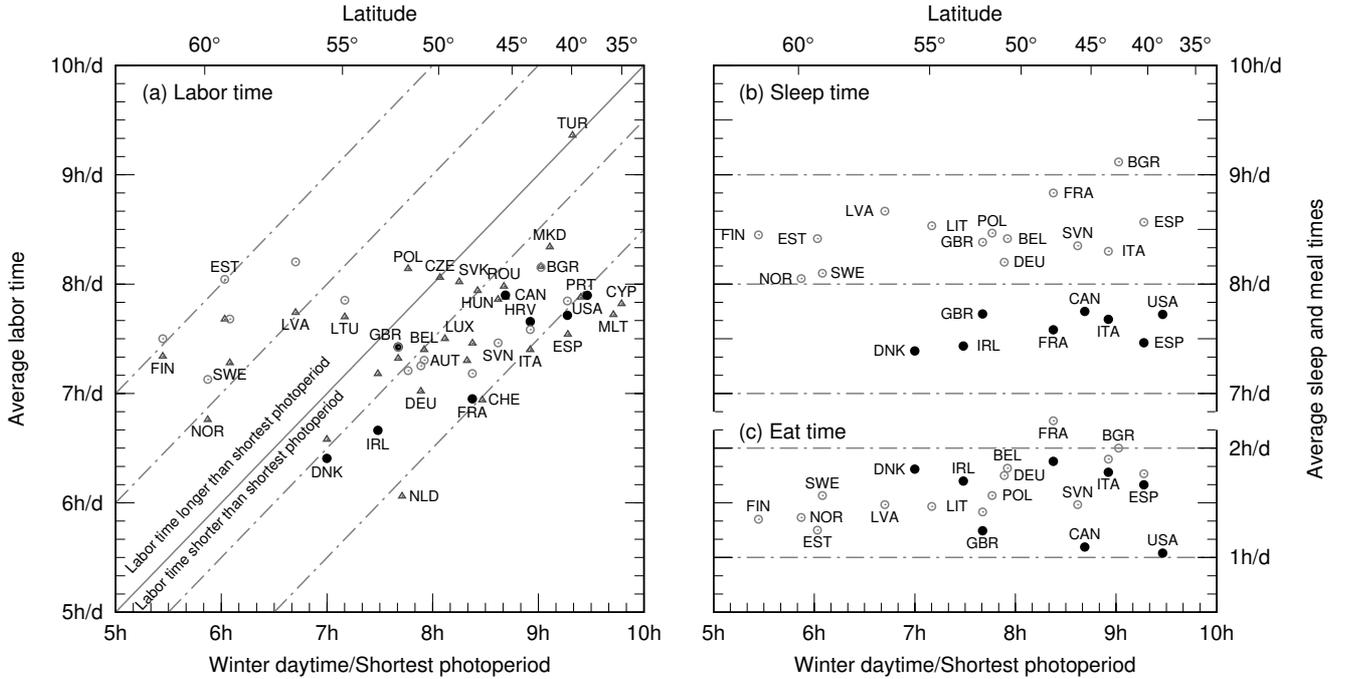}

  \caption{Average daily labor time (left), average daily sleep time (right top) and average daily eat time (right bottom)  versus winter daytime. Data from national time use surveys (solid circles), Hetus pre-prepared tables (open circles) and, on left panel, Eurostat (triangles). Two one-hour slanted strips are shown on the left panel: one shows a daily labor consumption \SI{1.5}{\hour} longer than the shortest photoperiod; the second one shows a daily labor consumption \SI{1}{\hour} shorter than the shortest photoperiod. In the right panels four horizontal strips are shown. Labels (only one for each country in panel (a) and (c)) indicate ISO-3166-1 alpha-3 country codes. Data are listed in Table~\ref{tab:duracion} and bivariate correlations are reported in Table~\ref{tab:ajusteDuracion}.}
  \label{fig:laborConsump}
\end{figure*}

Figure~\ref{fig:laborConsump} shows average daily consumptions for primary activities versus $D_w$ (bottom axis) or $\phi$ (top). Vertical axes show fractions of Earth's rotation period with $\SI{8}{\hour\per\day}\equiv1/3$ and $\SI{6}{\hour\per\day}\equiv1/4$.  Average labor time and average sleep time roughly account for one third of a day each.

In panel (b) sleep time from national time use surveys (solid circles) and Hetus pre-prepared tables (open circles) are shown. Although Hetus pre-prepared table reports the combined daily rhythm of ``sleeping and other personal care'' as discussed previously, it separately lists daily sleep time and other personal care daily time. The latter lies in the range of $\SI{40}{\min}$, a $\SI{7}{\percent}$ of the combined daily time.

Panel (b) shows the differences between standard population (Hetus, open symbols) and laborers (Time Use Surveys, solid symbols) which amounts to one whole hour less of daily sleep times. Table~\ref{tab:ajusteDuracion} shows a weak trend ($p=\SI{5.8(32)}{\min\per\hour}\sim\SI{-1}{\min\per\degree}; r^2=\num{0.197}$) with variability $2\us=\SI{20}{\min}$ for Hetus data. Variability is $\SI{7}{\percent}$ of the reference value. Employees in a week day yields similar figures but one hour less of sleep time and smaller variability ($\SI{4}{\percent}$ of $y_{\text{ref}}$). The lack of a significant latitudinal trend is understandable on the basis that sleep time is an indoor activity and that sleep time is likely a human universal.

In panel (c) daily eating consumption is presented. In this panel data sets stack vertically in the plot for a given country, whose label is displayed only once. Eat times lie in the range of one to two hours ---a tiny fraction of one day--- with remarkable latitudinal trend. For Hetus data $p=\SI{9.5(25)}{\min\per\hour}\sim\SI{-1.75}{\min\per\degree}; r^2=\num{0.513}$) with eat times decreasing with decreasing photoperiod. Hetus eat time variability is $\delta y=\SI{35}{\min}$ which is a $\SI{35}{\percent}$ of the daily eat time. Notwithstanding this, national time use surveys on laborers display opposite trend with meal times increasing with decreasing photoperiod due to the influence of the American data. Table~\ref{tab:ajusteDuracion} also reports bivariate correlations for daily TV time which show little correlation with latitude.
 
Panel (a) in Figure~\ref{fig:laborConsump} shows average daily labor consumption, which shapes the life of laborers. In this panel a third set of data has been included:  the average number of weekly hours of work\footnote{Weekly hours of work were converted into daily hours assuming a five day week} in main job obtained from Eurostat database\cite{eurostat-HoursWorked} (triangles) and referred to the year 2016. Again, data sets stack vertically and labels are displayed once. 

As well as panels (b) and (c), panel (a) in Figure~\ref{fig:laborConsump} can be read horizontally with most of data lying in the range of seven to eight hours a day, irrespective of latitude. Table~\ref{tab:ajusteDuracion} reports bivariate correlations for labor time and the subsets of data. It also reports  all of them combined.  Results vividly depend on data set or combination. Hetus data report lack of latitudinal dependence ($r^2<10^{-5}$) while national time use surveys soars to $r^2=\num{0.711}$ most likely due to small sample size and the influence of two of the countries with smallest labor time: France and Denmark. Eurostat data consists of $N=32$ countries (Iceland datum has not been considered) and finds correlations ($p=\SI{15.2(50)}{\min\per\hour}; r^2=\num{0.232}$) with $2\us=\SI{75}{\min}$ for the full range of latitudes, a $\SI{15}{\percent}$ of $y_{\text{ref}}=\text{7h35m}$.

Notwithstanding this, Figure~\ref{fig:laborConsump} shows slanted lines at a slope $\SI{60}{\min\per\hour}$ ---four times larger than slopes reported from bivariate correlations in Table~\ref{tab:ajusteDuracion}. Notice they are not deduced from data but highlight a one-to-one relationship between the shortest photoperiod and labor time. They provide an alternative insight: above \ang{54} latitude, daily labor consumption is \SIrange{1}{2}{\hour} longer than $D_w$; in contrast, below \ang{54} latitude, most data can be boxed in the range of \SIrange{30}{90}{\minute} shorter than $D_w$. Table~\ref{tab:ajusteDuracion} reports bivariate correlations for $\phi<\SI{54}{\degree}$ showing an increase in $r^2$ to $\num{0.338}$ and slope $p=\SI{28.8(64)}{\min\per\hour}$, which is half of the one-to-one coupling. It is remarkable considering the myriad of factors unrelated to light conditions that may punctuate labor time: productivity, gross domestic product or shares of economic sectors among others.

\section{Discussion}

\subsection{Should people living along a meridian be doing activities simultaneously?}
\label{sec:part-one}

Meridional properties characteristically occur at a given instant along a meridian  being solar noon the most remarkable of them. Mechanical clocks, conveniently tied to Earth's rotation, takes advantage of that and render time as a distance to noon, or to midnight.

Labor start time marks are not a meridional observation (see Figure~\ref{fig:labor} and Table~\ref{tab:ajustelabor}). It is likely that people living along a meridian over a wide range of latitude and different countries do not get to work simultaneously. Instead, year round, they will likely get to work at the latest sunrise time of the year, figure~\ref{fig:labor} show. To put it shortly, people living at $\SI{38}{\degree}$ may not feel the necessity of waiting until the Sun has risen at $\SI{55}{\degree}$ to start working. Correspondingly people at $\SI{55}{\degree}$ do not feel the necessity of start working once the Sun has already risen at $\SI{38}{\degree}$. 

Clock hours (7am/8am/9am) synchronize events like labor start but their significance ---too early, too late or fine--- is punctuated by light conditions ---distance to the terminator, not distance to noon--- and influenced by latitude. People make this translation and make rationale decisions accordingly populating the best choice. 

Likewise people living at $\SI{55}{\degree}$ are likely to agree leaving work earlier, agree coming home earlier or agree having dinning earlier than people living at $\SI{38}{\degree}$. All of these processes are likely synchronized by winter sunset time, specially in Europe.

The rationale of the winter synchronizer is:  once labor daily rhythm fits the worst case scenario, it only improves as the year progresses and the Sun apparently grows up. That way, and within some variability, labor daily rhythm is linked to ambient light conditions, with societies abhorring labor activity in darkness and harvesting insolation to produce goods.
 
Notice that the rationale for winter synchronization are hardly symmetrically transposed on summer. People living at $\SI{55}{\degree}$ seldom feel the necessity of start working earlier than people leaving at $\SI{38}{\degree}$ despite they will be observing earlier sunrise. Summer photoperiod $D_s\sim\SI{17}{\hour}$ at $\SI{55}{\degree}$ is exceedingly larger than average labor time. Summer time is not the worst case scenario. Likewise people living at lower latitudes are less likely to be driven by the winter sunrise time since it would increasingly exceed average labor time.

 As a general rule searching for latitudinal trends in time marks which span within one hour would require the span of latitudes be large enough so as some relevant time property ---shortest photoperiod--- spans by more than one hour.  At lunch time this issue is further complicated because competing synchronizers appear. People living along a meridian covering a wide latitude range are likely to have lunch simultaneously around noon (the first synchronizer). And they are also likely to break that behavior and have lunch three hours before winter sunset time (the second synchronizer), Figure~\ref{fig:meal} and Table~\ref{tab:ajusteComida} show. Competing synchronizers do not appear at labor start times or dinner times because there is no other significant natural event.

\subsection{Seasonality}
\label{sec:seasonality}

Seasonality is another issue which merits discussion.     Solar seasonality is large at middle latitudes: from winter to summer photoperiod changes by $\SI{6}{\hour}$ at $\phi\sim\SI{40}{\degree}$ and by $\SI{10}{\hour}$ at $\phi\sim\SI{55}{\degree}$. It is also symmetric, offsetting latitudinal trends to a yearly averaged photoperiod $\langle D\rangle=\SI{12}{\hour}$ irrespective of latitude. 
 
Figure~\ref{fig:activida} shows yearly averaged daily rhythms which soars in the morning. A measure to introduce is how fast activity is regained in the morning. For that purpose one may consider the elapsed time between the daily rhythm overshooting two convenient thresholds: \SI{25}{\percent}  and \SI{75}{\percent} of daily maximum activity. It shows an average value for labor activity  $\langle\Delta\tau\rangle=\SI{80}{\min}$ with $2s=\SI{30}{\min}$. Understandably similar figures can be obtained for the sleep-wake daily rhythm of laborers. For standard population  $\langle\Delta\tau\rangle=\SI{115}{\min}$ with $2s=\SI{20}{\min}$. 

The switch-off of the labor activity can not be analyzed in this way because it is a more complicated phenomenon which widens in time. As an example laborers who stop doing active work can start doing passive work: providing jobs opportunities for some other laborers; a process which unlikely happens in the morning. On the contrary the switch-off of the sleep-wake cycle can be analyzed this way yielding $\langle\Delta\tau\rangle=\SI{90}{\min}; 2s=\SI{30}{\min}$, irrespective of laborers or standard population.

There are two sources for $\Delta\tau$: (1) a seasonal distribution of rising times and labor start times: earlier times in summer compared to those of winter; and (2) a daily distribution of these times: people would unlikely rise and get to work at the same time on a daily basis. Since $\langle\Delta\tau\rangle$ is in the order of one hour option (2) looks reasonable and seasonality is weak compared to solar seasonality. 

Whereas in ancient times seasonality in human activity should be expected to some extend, it lost significance as mechanical clock ---which show no seasonality by itself--- started driving human activity, industrialization lead to decreased labor times and year round time schedules gained popularity. Today, Daylight Saving Times (DST) is likely the main, if not unique, source in seasonality.  

Notice that DST breaks the seasonal symmetry of the solar activity in a very specific way: sunrise local time variance decreases while sunset local time variance increases. That way DST inhibits morning seasonality: the appetence of laborers for earlier time schedules in summer. The important point is that societies would have likely abhorred breaking the symmetry the other way around ---narrowing the span of sunset times, enlarging the span of sunrise times--- because sunrise time is the prime synchronizer; not sunset times and not noon time. 

The calendar date when DST is set on and off also points towards the synchronizing role of winter sunrise. These dates depart from winter solstice date way enough so that the winter sunrise local time still be the latest year round and prevent labor start times from occurring before sunrise. DST was first used in 1916 during the Great War and then in the interwar period. Therefore, by then, time schedules may have been already arranged in a globally similar way to contemporary's: start times close to winter sunrise time and lack of seasonality. 

The survival of DST to the following winter puts time schedules into a major stress because winter sunrise time is delayed by one hour local time ---alternatively winter time schedules are advanced by one hour---. The important decisions that were rationally based on winter sunrise clock-time, now become irrational. It was not until 1938, during Spanish Civil War, and soon later in 1940 during World War Two, that DST survived to the following winter in many belligerent and non-belligerent countries. The survival of DST to the following winter is nowadays termed \emph{time zone advance}. 

War economy promotes time zone advance which, during peace time, often fails to prevail.  Examples of contemporary societies that could not sustain time zone advance for too long are: United Kingdom (1969-1971), Portugal (1967-1976 and 1992-1996), Russia (2011-2014) and Chile (2015). Darkness at labor start time is often the issue.

In contrast, time zone advance in the Canadian province of Saskatchewan (1960), Iceland (1969) and Alaska (1983) succeeded: people traded dark winter labor start times with lightful winter labor end times. That has also been the motivation of the recent (2017) time zone advance in the Chilean region of Magallanes. In every of these four regions absolute latitude is above \ang{50}. 

The Chilean case instructs the role of latitude in this issue. Noon occurs almost simultaneously everywhere in Chile; however it takes two hours for the winter and summer terminator to sweep the country. As a result  at the polar region of Magallanes both labor start and end times may likely occur in the winter darkness with $z_w<\ang{-6}$ if labor times do not change regionally in Chile. In May 2017 the time zone offset for this region effectively shifted from $\Delta=-\SI{4}{\hour}$ to $\Delta=-\SI{3}{\hour}$ so that labor winter end times will tend to occur before winter sunset or during twilight. The rationale for this time zone split is the winter terminator (latitude), instead of noon (longitude). 

France, Belgium and Spain, where time zone kept advanced at the end of World War Two (see $\delta>\SI{30}{\min}$ in Table~\ref{tab:geo}), fall in another category. In the wake of this turmoil the change survived simply because population quickly offset the advancing of time zone by delaying time schedules as historical records can track.\cite{kron-french-1901,bonilla-1907} This trade is more evident in Spain because of its Southwestern most location and can be observed in Tables~\ref{tab:data} to~\ref{tab:casa} by late local times. Mathematically they traded $t$ in Equation~(\ref{eq:6}) after $\Delta$ changed in a way that $\tau$ remained. Although this poses no harm for the population it jeopardizes the comparison of time schedules: if local times are to be compared, one whole hour should be subtracted to clock readings in these three countries ---or in any other region with advanced time zone--- as a rule of thumb. As an example morningness-eveningness tests can lean toward the eveningness\cite{Adan2002,Caci2005} if this rule of thumb is not taken into account. Strictly speaking this is a ``clock-time'' eveningness, and not a light/dark preference.

\subsection{Synchronization overturning}
\label{sec:synchr-overt}

The shortest photoperiod sets a significant non-meridional unit of time for life: people at $\SI{38}{\degree}$ latitude never live a photoperiod shorter than 9h30m as people at $\SI{55}{\degree}$ ($D_w\sim\SI{7}{\hour}$) latitude do. That makes a difference and influences the decision-making process that shapes human activity, notably the labor daily rhythm.

Figure~\ref{fig:laborConsump} (panel (a)) shows working population above $\SI{54}{\degree}$  must get used to the idea of doing activities in the winter darkness ($z_w<\SI{-6}{\degree}$) as observed in Figure~\ref{fig:labor}. On the contrary, below $\SI{54}{\degree}$ working population can accommodate their duties to winter daytime. Adding a reasonable lunch break, laborers struggle with darkness only in winter labor start time or in winter labor end time, but not both (see Figure~\ref{fig:labor} and $z_w$ in Table~\ref{tab:data}).

The far end of the strip highlights an average daily labor consumption $\SI{1.5}{\hour}$ shorter than winter daytime. In this scenario individuals and societies can find more ways to accommodate duties to light conditions.

It should be now stressed that there should exist a point during the photoperiod that turns the winter sunrise trend characterizing morning activity into the winter sunset trend, which characterizes evening activities. This is synchronization overturning. For the labor daily rhythm it happens in the afternoon when light conditions are steady and comfortable. Another overturning should occur at late night when light conditions are again steady.

Labor noon time marks (see Figure~\ref{fig:labor}) provides a rationale for this idea: data below $\SI{54}{\degree}$ lie in a one-hour strip centered at 4h30m after winter sunrise (see also Tables~\ref{tab:data} and~\ref{tab:ajustelabor}). Since every labor daily rhythms shows at dawn a fundamental rest state of least activity, by the time of labor noon cultural, social or even political differences have not yet grown enough to break the morning trend. Morning rhythms are similar from country to country and therefore it is afternoon rhythms that must proceed in fairly different ways from country to country so as to turn the trend upside down. Nature provides a clue: in winter, noon comes later in the morning as latitude decreases; yet, sunset is still further apart. 

These natural conditions may promote differences in decisions across countries and individuals: morning, afternoon and split shifts as an example. It also accounts for the variability in meal times (see Figure~\ref{fig:meal} and Table~\ref{tab:meal}) and lunch breaks. The excess of shortest photoperiod with respect to labor time play a significant role. 

 The position of the relative maximum in the labor daily rhythm observed after lunch break (see Figure~\ref{fig:activida} panel (c)) show by any kind of measure ---local time, mean solar time or distance to winter sunset---  a large variability. However a meaningful constraint can be put forward: the peak of activity after lunch break happens at least one hour before winter sunset, irrespective of latitude. Hence, labor activity never increases after sunset. It can be said in a statistical sense that laborers find comfortable extending lunch breaks as long as they do not return to work in the darkness.

Another example is the shares of laborers returning to home at lunch time, which can be tracked by  national time use surveys. Figure~\ref{fig:activida} panel (b) shows a decrease in the number of laborers out of home in Italy from \SIrange{93}{63}{\percent} at lunch time. A similar behavior is noted in Spain (\SIrange{90}{55}{\percent}) and France (\SIrange{82}{66}{\percent}). On the contrary, shares of laborers out of home in Great Britain, Denmark, Canada and United States keeps steady within a variance of $\SI{5}{\percent}$ or below.

 However, despite the variety of behaviors that can be observed during region of permanent light, societies find remarkably similar ways of activating and deactivating synchronized by the winter terminator.

Another synchronization overturning is likely to happen at late night. It should be noted that labor noon times occur some two hours before wakeful noon times (see $y_{\text{ref}}$ in Tables~\ref{tab:ajustelabor} and \ref{tab:ajusteVigil}). Hence labor activity leans to ``morning'' wakeful with ``afternoon'' wakeful prone to non-working activities. Evening activities are tied to the winter sunset and should sooner or later anticipate morning labor activities synchronized to the winter sunrise. The first example is TV prime time marks which likely occur meridionally (see Table~\ref{tab:ajusteTV}) so that two individuals living on the same meridian and different latitudes are likely to be watching TV simultaneously. 

An answer for sleeping times is less definite due to the differences in data for standard population and laborers. The former set could be less prone to light/dark cycle and hence its sleep-wake cycle less prone to latitudinal trends as observed (see Table~\ref{tab:ajusteVigil} and Figure~\ref{fig:vigil}). On the other hand Mid Sleep on Free days (MSF) ---the half-way point between sleep onset and sleep end, close in meaning to wakeful noon reported here for time use surveys--- a chronotype for standard population was reported\cite{Roenneberg2007} to correlate with summer sunrise time ---hence it would also correlate to winter sunset time--- for people in Germany.  However wake-up times for laborer should follow the signature as of the winter sunrise time which drives labor start times. 

Notwithstanding this, the remarkable coincidence of laborer and standard population bedtimes suggests that they might be distributed irrespective of latitude (see Figure~\ref{fig:vigil} and Table~\ref{tab:ajusteVigil}). The synchronizer overturning would make people living at lower latitude advance their bedtime ---despite they would have come home, have had dinner later--- because they should get up earlier. Correspondingly people living at higher latitudes may delay their bedtimes despite they would have arrived home, have had dinner earlier: they will get up later. Both behavior would result in the meridional, overturning behavior.

\section{Conclusion}
\label{sec:conclusion}

This work shows that year round time schedules have set the winter day ---the day with shortest photoperiod and with roughly the latest sunrise and earliest sunset--- as a source of synchronization for labor daily rhythm. It is the worst case scenario year round to which society finds accommodation. 

The winter day also influences laborers' rise times and meal times, especially in Europe. Variability related to cultural or inherited habits is still largely present as differences between American and European data show. This variability may be linked to latitude in the sense that the longer the shortest photoperiod the more pathways societies can test.

 Since the shortest photoperiod is determined by latitude, this property play a role in understanding social time albeit it is alien for clocks. Hence it is unlikely that a time mark in the morning, 8am for instance, would play the same role along a meridian because depending of latitude and calendar date it may mean whether the Sun has already risen or not. The same applies in the evening but not at forenoon, noon and afternoon.

This work shows that the winter sunrise, triggers the decision-making process by which societies within the range \SIrange{38}{54}{\degree} quickly come from the background state of rest in the darkness to the activity during photoperiod. Laborers get up, leave home, get to work quickly in its neighborhood year round, with DST as the only significant source of seasonality. While weak correlations can be found the main evidence is linked to placement: labor start times occur within the winter civil twilight.

Winter sunrise synchronization is overturned in the afternoon as the earliest sunset of the year, the winter sunset, gains significance. Data suggest that the winter sunset, trigger the opposite process in which individuals start making decisions to shut down labor activity.  Time use survey data can then track employees leaving work close to winter sunset time, returning to home some two hours after winter sunset, and then having dinner three hours after winter sunset time. Contrastingly, available data shows that TV prime times are irrespective of latitude suggesting a synchronization overturning at late night. 

The analysis of sleep-wake cycles does not lead to definite conclusions. Sleeping time of laborers seem to be understandably linked to winter sunrise time while sleep-wake cycle of standard population seems to be irrespective of latitude. It is an open question for future analysis if laborers' bedtimes could be meridional.

\section{Acknowledgments}
The author thanks the  institutions listed in References for providing access to national time use survey's microdata. Also to Statistics Finland for harmonizing the database of European time surveys and Statistics Sweden for building and running the table generating tool and making them publicly available. 

The author thanks Dr. Jos\'e Fern\'andez-Albertos at Centro de Ciencias Humanas y Sociales from Spanish CSIC for showing him the availability of time use surveys data and fruitful discussions thereafter.

The author is in debt with Prof. Dr. Jorge Mira P\'erez at Universidade de Santiago de Compostela for fruitful and encouraging discussions.

The author thanks to Spanish think tank Politikon (\url{http://www.politikon.es}) and to his editor Octavio Medina for their help disseminating the ideas explained in this paper.

The author also thanks to Captain (R) Juan Palacio now retired, formerly at The Royal Institute and Observatory of the Navy (ROA) in San Fernando for fruitful discussions and to Ms Ana Vega (twitter user @biscayenne) for her help with historical meal habits.



\begin{thebibliography}{1}
\expandafter\ifx\csname natexlab\endcsname\relax\def\natexlab#1{#1}\fi
\expandafter\ifx\csname bibnamefont\endcsname\relax
  \def\bibnamefont#1{#1}\fi
\expandafter\ifx\csname bibfnamefont\endcsname\relax
  \def\bibfnamefont#1{#1}\fi
\expandafter\ifx\csname citenamefont\endcsname\relax
  \def\citenamefont#1{#1}\fi
\expandafter\ifx\csname url\endcsname\relax
  \def\url#1{\texttt{#1}}\fi
\expandafter\ifx\csname urlprefix\endcsname\relax\def\urlprefix{URL }\fi
\providecommand{\bibinfo}[2]{#2}
\providecommand{\eprint}[2][]{\url{#2}}

\bibitem[{\citenamefont{Phillips}(2009)}]{Phillips2009}
\bibinfo{author}{\bibfnamefont{M.~L.} \bibnamefont{Phillips}},
  \bibinfo{journal}{Nature} \textbf{\bibinfo{volume}{458}},
  \bibinfo{pages}{142} (\bibinfo{year}{2009}), ISSN \bibinfo{issn}{0028-0836},
  \urlprefix\url{http://www.nature.com/doifinder/10.1038/458142a}.

\bibitem[{\citenamefont{Kramer and Merrow}(2013)}]{Kramer2013}
\bibinfo{author}{\bibfnamefont{A.}~\bibnamefont{Kramer}} \bibnamefont{and}
  \bibinfo{author}{\bibfnamefont{M.~e.} \bibnamefont{Merrow}},
  \emph{\bibinfo{title}{{Circadian Clocks}}}, vol. \bibinfo{volume}{217}
  (\bibinfo{publisher}{Springer}, \bibinfo{address}{Heidelberg},
  \bibinfo{year}{2013}),
  \urlprefix\url{http://www.springer.com/us/book/9783642259494}.

\bibitem[{imc(1884)}]{imc-1884}
\emph{\bibinfo{title}{Protocols of Proceedings of the {I}nternational
  {C}onference for the purpose of fixing a {P}rime {M}eridian and a universal
  day}} (\bibinfo{publisher}{Gibson Bros.}, \bibinfo{year}{1884}),
  \urlprefix\url{http://bit.ly/2tX2pQi}.

\bibitem[{\citenamefont{Roenneberg et~al.}(2013)\citenamefont{Roenneberg,
  Kantermann, Juda, Vetter, and Allebrandt}}]{Roenneberg2013}
\bibinfo{author}{\bibfnamefont{T.}~\bibnamefont{Roenneberg}},
  \bibinfo{author}{\bibfnamefont{T.}~\bibnamefont{Kantermann}},
  \bibinfo{author}{\bibfnamefont{M.}~\bibnamefont{Juda}},
  \bibinfo{author}{\bibfnamefont{C.}~\bibnamefont{Vetter}}, \bibnamefont{and}
  \bibinfo{author}{\bibfnamefont{K.~V.} \bibnamefont{Allebrandt}}, in
  \emph{\bibinfo{booktitle}{Circadian Clocks}} (\bibinfo{publisher}{Springer,
  Berlin, Heidelberg}, \bibinfo{year}{2013}), pp. \bibinfo{pages}{311--331}.

\bibitem[{\citenamefont{Horne and Ostberg}(1976)}]{Horne1976}
\bibinfo{author}{\bibfnamefont{J.~A.} \bibnamefont{Horne}} \bibnamefont{and}
  \bibinfo{author}{\bibfnamefont{O.}~\bibnamefont{Ostberg}},
  \bibinfo{journal}{Int. J. Chronobiol.} \textbf{\bibinfo{volume}{4}},
  \bibinfo{pages}{97} (\bibinfo{year}{1976}), ISSN \bibinfo{issn}{0300-9998},
  \urlprefix\url{http://www.ncbi.nlm.nih.gov/pubmed/1027738}.

\bibitem[{\citenamefont{Smith et~al.}(1989)\citenamefont{Smith, Reilly, and
  Midkiff}}]{Smith1989}
\bibinfo{author}{\bibfnamefont{C.~S.} \bibnamefont{Smith}},
  \bibinfo{author}{\bibfnamefont{C.}~\bibnamefont{Reilly}}, \bibnamefont{and}
  \bibinfo{author}{\bibfnamefont{K.}~\bibnamefont{Midkiff}},
  \bibinfo{journal}{J. Appl. Psychol.} \textbf{\bibinfo{volume}{74}},
  \bibinfo{pages}{728} (\bibinfo{year}{1989}), ISSN \bibinfo{issn}{1939-1854},
  \urlprefix\url{http://doi.apa.org/getdoi.cfm?doi=10.1037/0021-9010.74.5.728}.

\bibitem[{\citenamefont{Roenneberg et~al.}(2003)\citenamefont{Roenneberg,
  Wirz-Justice, and Merrow}}]{Roenneberg2003}
\bibinfo{author}{\bibfnamefont{T.}~\bibnamefont{Roenneberg}},
  \bibinfo{author}{\bibfnamefont{A.}~\bibnamefont{Wirz-Justice}},
  \bibnamefont{and} \bibinfo{author}{\bibfnamefont{M.}~\bibnamefont{Merrow}},
  \bibinfo{journal}{J. Biol. Rhythms} \textbf{\bibinfo{volume}{18}},
  \bibinfo{pages}{80} (\bibinfo{year}{2003}), ISSN \bibinfo{issn}{0748-7304},
  \urlprefix\url{http://journals.sagepub.com/doi/10.1177/0748730402239679}.

\bibitem[{\citenamefont{White and Terman}(2003)}]{White2003}
\bibinfo{author}{\bibfnamefont{T.~M.} \bibnamefont{White}} \bibnamefont{and}
  \bibinfo{author}{\bibfnamefont{M.}~\bibnamefont{Terman}},
  \bibinfo{journal}{Psychiatry Interpers. Biol. Process.}
  \textbf{\bibinfo{volume}{742}}, \bibinfo{pages}{1193} (\bibinfo{year}{2003}).

\bibitem[{\citenamefont{Roenneberg et~al.}(2007)\citenamefont{Roenneberg,
  Kumar, and Merrow}}]{Roenneberg2007}
\bibinfo{author}{\bibfnamefont{T.}~\bibnamefont{Roenneberg}},
  \bibinfo{author}{\bibfnamefont{C.~J.} \bibnamefont{Kumar}}, \bibnamefont{and}
  \bibinfo{author}{\bibfnamefont{M.}~\bibnamefont{Merrow}},
  \bibinfo{journal}{Curr. Biol.} \textbf{\bibinfo{volume}{17}},
  \bibinfo{pages}{R44} (\bibinfo{year}{2007}), ISSN \bibinfo{issn}{09609822},
  \urlprefix\url{http://linkinghub.elsevier.com/retrieve/pii/S0960982206026091}.

\bibitem[{\citenamefont{Randler}(2008)}]{Randler2008}
\bibinfo{author}{\bibfnamefont{C.}~\bibnamefont{Randler}},
  \bibinfo{journal}{Chronobiol. Int.} \textbf{\bibinfo{volume}{25}},
  \bibinfo{pages}{1017} (\bibinfo{year}{2008}), ISSN \bibinfo{issn}{0742-0528},
  \urlprefix\url{http://www.tandfonline.com/doi/full/10.1080/07420520802551519}.

\bibitem[{\citenamefont{Miguel et~al.}(2014)\citenamefont{Miguel, de~Oliveira,
  Pereira, and Pedrazzoli}}]{Miguel2014a}
\bibinfo{author}{\bibfnamefont{M.}~\bibnamefont{Miguel}},
  \bibinfo{author}{\bibfnamefont{V.~C.} \bibnamefont{de~Oliveira}},
  \bibinfo{author}{\bibfnamefont{D.}~\bibnamefont{Pereira}}, \bibnamefont{and}
  \bibinfo{author}{\bibfnamefont{M.}~\bibnamefont{Pedrazzoli}},
  \bibinfo{journal}{Ann. Hum. Biol.} \textbf{\bibinfo{volume}{41}},
  \bibinfo{pages}{107} (\bibinfo{year}{2014}), ISSN \bibinfo{issn}{0301-4460},
  \urlprefix\url{http://www.ncbi.nlm.nih.gov/pubmed/24059265
  http://www.tandfonline.com/doi/full/10.3109/03014460.2013.832795}.

\bibitem[{\citenamefont{Leocadio-Miguel
  et~al.}(2017)\citenamefont{Leocadio-Miguel, Louzada, Duarte, Areas, Alam,
  Freire, Fontenele-Araujo, Menna-Barreto, and
  Pedrazzoli}}]{Leocadio-Miguel2017}
\bibinfo{author}{\bibfnamefont{M.~A.} \bibnamefont{Leocadio-Miguel}},
  \bibinfo{author}{\bibfnamefont{F.~M.} \bibnamefont{Louzada}},
  \bibinfo{author}{\bibfnamefont{L.~L.} \bibnamefont{Duarte}},
  \bibinfo{author}{\bibfnamefont{R.~P.} \bibnamefont{Areas}},
  \bibinfo{author}{\bibfnamefont{M.}~\bibnamefont{Alam}},
  \bibinfo{author}{\bibfnamefont{M.~V.} \bibnamefont{Freire}},
  \bibinfo{author}{\bibfnamefont{J.}~\bibnamefont{Fontenele-Araujo}},
  \bibinfo{author}{\bibfnamefont{L.}~\bibnamefont{Menna-Barreto}},
  \bibnamefont{and}
  \bibinfo{author}{\bibfnamefont{M.}~\bibnamefont{Pedrazzoli}},
  \bibinfo{journal}{Sci. Rep.} \textbf{\bibinfo{volume}{7}},
  \bibinfo{pages}{5437} (\bibinfo{year}{2017}),
  \urlprefix\url{https://www.nature.com/articles/s41598-017-05797-w}.

\bibitem[{\citenamefont{Randler and Rahafar}(2017)}]{Randler2017}
\bibinfo{author}{\bibfnamefont{C.}~\bibnamefont{Randler}} \bibnamefont{and}
  \bibinfo{author}{\bibfnamefont{A.}~\bibnamefont{Rahafar}},
  \bibinfo{journal}{Sci. Rep.} \textbf{\bibinfo{volume}{7}},
  \bibinfo{pages}{39976} (\bibinfo{year}{2017}), ISSN
  \bibinfo{issn}{2045-2322},
  \urlprefix\url{http://www.nature.com/articles/srep39976}.

\bibitem[{\citenamefont{Monsivais et~al.}(2017)\citenamefont{Monsivais, Ghosh,
  Bhattacharya, Dunbar, and Kaski}}]{Monsivais2017}
\bibinfo{author}{\bibfnamefont{D.}~\bibnamefont{Monsivais}},
  \bibinfo{author}{\bibfnamefont{A.}~\bibnamefont{Ghosh}},
  \bibinfo{author}{\bibfnamefont{K.}~\bibnamefont{Bhattacharya}},
  \bibinfo{author}{\bibfnamefont{R.~I.~M.} \bibnamefont{Dunbar}},
  \bibnamefont{and} \bibinfo{author}{\bibfnamefont{K.}~\bibnamefont{Kaski}},
  \bibinfo{journal}{PLOS Comput. Biol.} \textbf{\bibinfo{volume}{13}},
  \bibinfo{pages}{e1005824} (\bibinfo{year}{2017}), ISSN
  \bibinfo{issn}{1553-7358},
  \urlprefix\url{http://dx.plos.org/10.1371/journal.pcbi.1005824}.

\bibitem[{\citenamefont{Chenu and Lesnard}(2006)}]{Chenu2006}
\bibinfo{author}{\bibfnamefont{A.}~\bibnamefont{Chenu}} \bibnamefont{and}
  \bibinfo{author}{\bibfnamefont{L.}~\bibnamefont{Lesnard}},
  \bibinfo{journal}{Eur. J. Sociol.} \textbf{\bibinfo{volume}{47}},
  \bibinfo{pages}{335} (\bibinfo{year}{2006}), ISSN \bibinfo{issn}{0003-9756},
  \urlprefix\url{http://www.journals.cambridge.org/abstract{\_}S0003975606000117}.

\bibitem[{\citenamefont{Craig}(2006)}]{craig2006}
\bibinfo{author}{\bibfnamefont{L.}~\bibnamefont{Craig}},
  \bibinfo{journal}{Gend. Soc.} \textbf{\bibinfo{volume}{20}},
  \bibinfo{pages}{259} (\bibinfo{year}{2006}),
  \urlprefix\url{http://journals.sagepub.com/doi/pdf/10.1177/0891243205285212}.

\bibitem[{\citenamefont{Aguiar and Hurst}(2007)}]{aguiar2007}
\bibinfo{author}{\bibfnamefont{M.}~\bibnamefont{Aguiar}} \bibnamefont{and}
  \bibinfo{author}{\bibfnamefont{E.}~\bibnamefont{Hurst}}, \bibinfo{journal}{Q.
  J. Econ.} \textbf{\bibinfo{volume}{122}}, \bibinfo{pages}{969}
  (\bibinfo{year}{2007}), ISSN \bibinfo{issn}{0033-5533},
  \urlprefix\url{http://dx.doi.org/10.1162/qjec.122.3.969}.

\bibitem[{\citenamefont{Aguiar et~al.}(2013)\citenamefont{Aguiar, Hurst, and
  Karabarbounis}}]{aguiar2013}
\bibinfo{author}{\bibfnamefont{M.}~\bibnamefont{Aguiar}},
  \bibinfo{author}{\bibfnamefont{E.}~\bibnamefont{Hurst}}, \bibnamefont{and}
  \bibinfo{author}{\bibfnamefont{L.}~\bibnamefont{Karabarbounis}},
  \bibinfo{journal}{Am. Econ. Rev.} \textbf{\bibinfo{volume}{103}},
  \bibinfo{pages}{1664} (\bibinfo{year}{2013}),
  \urlprefix\url{http://www.aeaweb.org/articles?id=10.1257/aer.103.5.1664}.

\bibitem[{\citenamefont{{Det Nationale Forkskningscenter for Velf\ae
  rd}}(2001)}]{dktus-2001e}
\bibinfo{author}{\bibnamefont{{Det Nationale Forkskningscenter for Velf\ae
  rd}}}, \emph{\bibinfo{title}{{D}anish {T}ime {U}se {S}urvey: {D}anske
  {T}idsanvendelseunders{\o}gelsen}}, \bibinfo{howpublished}{{C}enter for
  {S}urvey and {S}urvey/{R}egister {D}ata (distribuitor)}
  (\bibinfo{year}{2001}), \urlprefix\url{http://bit.ly/2t37Ksj}.

\bibitem[{\citenamefont{{Instituto Nacional de
  Estadística}}(2010)}]{estus-2010e}
\bibinfo{author}{\bibnamefont{{Instituto Nacional de Estadística}}},
  \emph{\bibinfo{title}{{S}panish {T}ime {U}se {S}urvey: {E}ncuesta de {E}mpleo
  del {T}iempo}} (\bibinfo{year}{2010}), \urlprefix\url{http://bit.ly/2uQFqa0}.

\bibitem[{\citenamefont{{Bureau of Labor Statistics}}(2012)}]{ustus-2012}
\bibinfo{author}{\bibnamefont{{Bureau of Labor Statistics}}},
  \emph{\bibinfo{title}{{A}merican {T}ime {U}se {S}urvey}},
  \bibinfo{howpublished}{computer file (multi year data)}
  (\bibinfo{year}{2012}),
  \urlprefix\url{https://www.bls.gov/tus/datafiles_2013.htm}.

\bibitem[{\citenamefont{{L'Institut National de la Statisque et des Études
  Économiques}}(2010)}]{frtus-2010e}
\bibinfo{author}{\bibnamefont{{L'Institut National de la Statisque et des
  Études Économiques}}}, \emph{\bibinfo{title}{{F}rench {T}ime {U}se
  {S}urvey. {E}nqu\^ete emploi du {T}emps et {D}écisions dans les couples}}
  (\bibinfo{year}{2010}).

\bibitem[{\citenamefont{{Economic and Social Research
  Institute}}()}]{ietus-2005}
\bibinfo{author}{\bibnamefont{{Economic and Social Research Institute}}},
  \emph{\bibinfo{title}{The {I}rish {N}ational {T}ime-{U}se {S}urvey}},
  \bibinfo{howpublished}{{I}rish {S}ocial {S}ience {D}ata {A}rchive
  (distribuitor)}.

\bibitem[{\citenamefont{{L'Istituto nazionale di statistica
  (Istat)}}(2009)}]{ittus-2010e}
\bibinfo{author}{\bibnamefont{{L'Istituto nazionale di statistica (Istat)}}},
  \emph{\bibinfo{title}{{I}talian {T}ime {U}se {S}urvey: Uso del tempo}}
  (\bibinfo{year}{2009}).

\bibitem[{\citenamefont{{Ipsos-RSL and Office of National
  Statistics}}(2003)}]{uktus-2003}
\bibinfo{author}{\bibnamefont{{Ipsos-RSL and Office of National Statistics}}},
  \emph{\bibinfo{title}{{U}nited {K}ingdom {T}ime {U}se {S}urvey 2000 (computer
  file)}}, \bibinfo{howpublished}{3rd ed, {C}olchester, {E}ssex: {UK} {D}ata
  archive (distribuitor)} (\bibinfo{year}{2003}).

\bibitem[{\citenamefont{{Statistics Canada/Statisque
  Canada}}(2005)}]{catus-2005}
\bibinfo{author}{\bibnamefont{{Statistics Canada/Statisque Canada}}},
  \emph{\bibinfo{title}{General {S}ocial {S}urvey, {T}ime {U}se. cycle 19}},
  \bibinfo{howpublished}{computer file} (\bibinfo{year}{2005}).

\bibitem[{\citenamefont{{Statistical Office of the European Commission
  (Eurostat)}}(2013)}]{eurostat-nuts}
\bibinfo{editor}{\bibnamefont{{Statistical Office of the European Commission
  (Eurostat)}}}, ed., \emph{\bibinfo{title}{Regions in the {E}uropean {U}nion.
  {N}omenclature of {T}erritorial {U}nits for {S}tatistics}}
  (\bibinfo{publisher}{Office for Official Publications of the European
  Communities}, \bibinfo{year}{2013}), ISBN \bibinfo{isbn}{978-92-79-38657-2},
  \urlprefix\url{http://bit.ly/2uq7Slz}.

\bibitem[{\citenamefont{{Statistical Office of the European Commission
  (Eurostat)}}(2009)}]{eurostat-hetus}
\bibinfo{editor}{\bibnamefont{{Statistical Office of the European Commission
  (Eurostat)}}}, ed., \emph{\bibinfo{title}{Harmonised {E}uropean time use
  surveys, 2008 guidelines}} (\bibinfo{publisher}{Office for Official
  Publications of the European Communities}, \bibinfo{year}{2009}), ISBN
  \bibinfo{isbn}{978-92-79-07853-8}, \urlprefix\url{http://bit.ly/2tXTIpb}.

\bibitem[{\citenamefont{{Statistics Finland} and {Statistics
  Sweden}}(2005-2007)}]{hetus}
\bibinfo{author}{\bibnamefont{{Statistics Finland}}} \bibnamefont{and}
  \bibinfo{author}{\bibnamefont{{Statistics Sweden}}},
  \emph{\bibinfo{title}{{H}armonised {E}uropean {T}ime {U}se {S}urvey [online
  database version 2.0]}} (\bibinfo{year}{2005-2007}),
  \urlprefix\url{http://www.tus.scb.se}.

\bibitem[{\citenamefont{Pierrehumbert}(2014)}]{pierre-cam-2014}
\bibinfo{author}{\bibfnamefont{R.~T.} \bibnamefont{Pierrehumbert}},
  \emph{\bibinfo{title}{Principles of Planetary Climate}}
  (\bibinfo{publisher}{Cambridge Univ. Press}, \bibinfo{year}{2014}), ISBN
  \bibinfo{isbn}{978-0521865562}, \urlprefix\url{http://bit.ly/1X7B9q8}.

\bibitem[{\citenamefont{{Statistical Office of the European Commission
  (Eurostat)}}(2017)}]{eurostat-HoursWorked}
\bibinfo{author}{\bibnamefont{{Statistical Office of the European Commission
  (Eurostat)}}}, \emph{\bibinfo{title}{Average number of usual weekly hours of
  work in main job [code: lfsa\_ewhun2]}} (\bibinfo{year}{2017}),
  \urlprefix\url{http://bit.ly/2ulnt5G}.

\bibitem[{\citenamefont{Kron}(1901)}]{kron-french-1901}
\bibinfo{author}{\bibfnamefont{R.}~\bibnamefont{Kron}},
  \emph{\bibinfo{title}{French Daily Life}} (\bibinfo{publisher}{Newson and
  Company}, \bibinfo{address}{New York}, \bibinfo{year}{1901}),
  \urlprefix\url{http://bit.ly/2lzIG5d}.

\bibitem[{\citenamefont{Bonilla}(1907)}]{bonilla-1907}
\bibinfo{author}{\bibfnamefont{R.~H.} \bibnamefont{Bonilla}},
  \emph{\bibinfo{title}{Spanish Daily Life}} (\bibinfo{publisher}{Newson and
  Company}, \bibinfo{address}{New York}, \bibinfo{year}{1907}),
  \urlprefix\url{http://bit.ly/2mocm4H}.

\bibitem[{\citenamefont{Adan and Natale}(2002)}]{Adan2002}
\bibinfo{author}{\bibfnamefont{A.}~\bibnamefont{Adan}} \bibnamefont{and}
  \bibinfo{author}{\bibfnamefont{V.}~\bibnamefont{Natale}},
  \bibinfo{journal}{Chronobiol. Int.} \textbf{\bibinfo{volume}{19}},
  \bibinfo{pages}{709} (\bibinfo{year}{2002}), ISSN \bibinfo{issn}{0742-0528},
  \urlprefix\url{http://www.ncbi.nlm.nih.gov/pubmed/12182498}.

\bibitem[{\citenamefont{Caci et~al.}(2005)\citenamefont{Caci, Adan, Bohle,
  Natale, Pornpitakpan, and Tilley}}]{Caci2005}
\bibinfo{author}{\bibfnamefont{H.}~\bibnamefont{Caci}},
  \bibinfo{author}{\bibfnamefont{A.}~\bibnamefont{Adan}},
  \bibinfo{author}{\bibfnamefont{P.}~\bibnamefont{Bohle}},
  \bibinfo{author}{\bibfnamefont{V.}~\bibnamefont{Natale}},
  \bibinfo{author}{\bibfnamefont{C.}~\bibnamefont{Pornpitakpan}},
  \bibnamefont{and} \bibinfo{author}{\bibfnamefont{A.}~\bibnamefont{Tilley}},
  \bibinfo{journal}{Chronobiol. Int.} \textbf{\bibinfo{volume}{22}},
  \bibinfo{pages}{523} (\bibinfo{year}{2005}), ISSN \bibinfo{issn}{0742-0528},
  \urlprefix\url{http://www.tandfonline.com/doi/full/10.1081/CBI-200062401}.

\end{thebibliography}


\appendix

\section{Tables of results}
\label{sec:table-results}

\begin{table*}
  \centering
\setlength{\tabcolsep}{2pt}

\begin{tabular}{llc|rr|ccc}
\toprule
\textbf{Country}&\textbf{Label}&\textbf{Set}&\textbf{Latitude}&\textbf{Time offset}&\textbf{Sunrise}&\textbf{Daytime}&\textbf{Sunset}\\
\multicolumn{1}{c}{}&\multicolumn{2}{c}{}&\multicolumn{1}{c}{$\phi$}&\multicolumn{1}{c}{$\delta=\Delta-\lambda/\Omega$}&\multicolumn{1}{c}{$t^{\uparrow}_w$}&\multicolumn{1}{c}{$D_w$}&\multicolumn{1}{c}{$t^{\downarrow}_w$}\\
\toprule
\multicolumn{8}{l}{\emph{Europe $\phi<\ang{54}$}}\\
\midrule
Spain&ESP&1,2&$\ang{40.4}$&$\SI{72}{min}$&08:34&09h17m&17:50\\
Bulgaria&BGR&2&$\ang{42.7}$&$\SI{21}{min}$&07:50&09h01m&16:52\\
Italy&ITA&1,2&$\ang{43.6}$&$\SI{11}{min}$&07:43&08h55m&16:39\\
Slovenia&SVN&2&$\ang{46.1}$&$\SI{ 2}{min}$&07:43&08h37m&16:21\\
France&FRA&1,2&$\ang{47.8}$&$\SI{50}{min}$&08:39&08h23m&17:01\\
Belgium&BEL&2&$\ang{50.9}$&$\SI{42}{min}$&08:45&07h55m&16:40\\
Germany&DEU&2&$\ang{51.0}$&$\SI{23}{min}$&08:26&07h53m&16:20\\
Poland&POL&2&$\ang{51.8}$&$\SI{-17}{min}$&07:50&07h46m&15:36\\
United Kingdom&GBR&1,2&$\ang{52.3}$&$\SI{ 6}{min}$&08:16&07h40m&15:56\\
Ireland&IRL&1&$\ang{53.3}$&$\SI{25}{min}$&08:41&07h29m&16:10\\
\midrule
\multicolumn{3}{l|}{Average}&$\ang{48.0}$&$\SI{24}{min}$&08:15&08h18m&16:32\\
\multicolumn{3}{l|}{Variability\hfill $2s(\{x_i\})$}&$\ang{9.1}$&$\SI{52}{min}$&$\SI{51}{\min}$&01h16m&01h15m\\
\toprule
\multicolumn{8}{l}{\emph{Europe $\phi>\ang{54}$}}\\
\midrule
Lithuania&LIT&2&$\ang{54.9}$&$\SI{24}{min}$&08:49&07h10m&15:59\\
Denmark&DNK&1&$\ang{55.7}$&$\SI{13}{min}$&08:43&07h00m&15:43\\
Latvia&LVA&2&$\ang{56.9}$&$\SI{24}{min}$&09:03&06h42m&15:45\\
Sweden&SWE&2&$\ang{59.2}$&$\SI{-5}{min}$&08:53&06h05m&14:58\\
Estonia&EST&2&$\ang{59.4}$&$\SI{21}{min}$&09:20&06h02m&15:22\\
Norway&NOR&2&$\ang{59.9}$&$\SI{18}{min}$&09:22&05h52m&15:14\\
Finland&FIN&2&$\ang{61.1}$&$\SI{20}{min}$&09:37&05h27m&15:04\\
\midrule
\multicolumn{3}{l|}{Average}&$\ang{58.2}$&$\SI{16}{min}$&09:07&06h20m&15:26\\
\multicolumn{3}{l|}{Variability\hfill $2s(\{x_i\})$}&$\ang{4.7}$&$\SI{20}{min}$&$\SI{40}{\min}$&01h16m&$\SI{46}{\min}$\\
\toprule
\multicolumn{8}{l}{\emph{Europe all}}\\
\midrule
\multicolumn{3}{l|}{Average}&$\ang{52.2}$&$\SI{21}{min}$&08:36&07h29m&16:05\\
\multicolumn{3}{l|}{Variability\hfill $2s(\{x_i\})$}&$\ang{12.7}$&$\SI{41}{min}$&01h10m&02h21m&01h32m\\
\toprule
\multicolumn{8}{l}{\emph{America}}\\
\midrule
United States&USA&1&$\ang{38.5}$&$\SI{ 8}{min}$&07:24&09h28m&16:52\\
Canada&CAN&1&$\ang{45.5}$&$\SI{17}{min}$&07:56&08h41m&16:38\\
\multicolumn{3}{l|}{Average}&$\ang{42.0}$&$\SI{12}{min}$&07:40&09h05m&16:45\\
\multicolumn{3}{l|}{Variability\hfill $2s(\{x_i\})$}&$\ang{9.9}$&$\SI{13}{min}$&$\SI{46}{\min}$&01h06m&$\SI{20}{\min}$\\
\toprule
\multicolumn{8}{l}{\emph{All data}}\\
\midrule
\multicolumn{3}{l|}{Average}&$\ang{51.1}$&$\SI{20}{min}$&08:30&07h39m&16:09\\
\multicolumn{3}{l|}{Variability\hfill $2s(\{x_i\})$}&$\ang{13.8}$&$\SI{39}{min}$&01h15m&02h26m&01h30m\\
\bottomrule
\end{tabular}

  \caption{Overview of geographical data for participating countries whose data were retrieve from national time use surveys (1) or Harmonised European Time Use Surveys (2). Latitude (rounded to one tenth of degree) and time offset (rounded to one minute) values are weighted population median values. Winter sunrise time $t^\uparrow_w$, winter daylength $D_w$ and winter sunset time $t^\downarrow_w$ where computed through Equations~(\ref{eq:2}),~(\ref{eq:9}), and~(\ref{eq:6}) using the listed values of latitude and time offset. Sunrise and sunset times are local times. Time zones are not listed. For each subset sample average value and the variability expressed as twice sample standard deviation are listed.}
  \label{tab:geo}
\end{table*}
\begin{table*}
  \centering
\setlength{\tabcolsep}{2pt}
\begin{tabular}{lc|ccr|ccc|ccr}
\toprule
\textbf{Country}&\textbf{Label}&\multicolumn{3}{c}{\textbf{Labor Start}}&\multicolumn{3}{c}{\textbf{Labor Noon}}&\multicolumn{3}{c}{\textbf{Labor End}}\\ 
&&$t$&$\tau$&$z_w$&$t$&$\tau$&$\Delta t_w$&$t$&$\tau$&$z_w$\\ 
\midrule
\multicolumn{11}{l}{\emph{Time Use Surveys (employees in weekday only)}}\\
\midrule
United States&USA&07:25&07:20&$\SI{00}{\degree}$&12:35&12:25&$+$05h10m&16:55&16:45&$\SI{-1}{\degree}$\\ 
Spain&ESP&08:05&06:55&$\SI{-6}{\degree}$&12:40&11:30&$+$04h10m&18:20&17:10&$\SI{-6}{\degree}$\\ 
Italy&ITA&08:00&07:45&$\SI{01}{\degree}$&12:20&12:10&$+$04h40m&17:45&17:35&$\SI{-12}{\degree}$\\ 
Canada&CAN&07:45&07:30&$\SI{-2}{\degree}$&12:55&12:40&$+$05h00m&16:55&16:40&$\SI{-4}{\degree}$\\ 
France&FRA&08:10&07:20&$\SI{-5}{\degree}$&13:20&12:30&$+$04h40m&17:40&16:50&$\SI{-6}{\degree}$\\ 
United Kingdom&GBR&08:15&08:10&$\SI{-1}{\degree}$&12:55&12:45&$+$04h40m&17:00&16:55&$\SI{-9}{\degree}$\\ 
Ireland&IRL&08:45&08:15&$\SI{-1}{\degree}$&13:30&13:00&$+$04h45m&17:15&16:45&$\SI{-9}{\degree}$\\ 
Denmark&DNK&07:50&07:40&$\SI{-7}{\degree}$&12:30&12:15&$+$03h45m&16:00&15:45&$\SI{-3}{\degree}$\\ 
\midrule
\multicolumn{2}{l|}{Average}&08:00&07:35&\SI{-3}{\degree}&12:50&12:25&04h35m&17:15&16:50&\SI{-6}{\degree}\\
\multicolumn{2}{l|}{Variability \hfill$2\us(\{x_i\})$}&\SI{45}{\min}&\SI{55}{\min}&\SI{+6}{\degree}&\SI{50}{\min}&\SI{55}{\min}&\SI{55}{\min}&01h25m&01h00m&\SI{+7}{\degree}\\
\bottomrule
\multicolumn{11}{l}{\emph{Hetus pre-prepared tables (standard population)}}\\
\midrule
Spain&ESP&08:10&07:00&$\SI{-5}{\degree}$&13:00&11:50&$+$04h25m&19:10&18:00&$\SI{-15}{\degree}$\\ 
Bulgaria&BGR&08:00&07:40&$\SI{01}{\degree}$&13:00&12:40&$+$05h10m&17:00&16:40&$\SI{-2}{\degree}$\\ 
Italy&ITA&08:00&07:50&$\SI{02}{\degree}$&12:30&12:20&$+$04h45m&18:00&17:50&$\SI{-14}{\degree}$\\ 
Slovenia&SVN&07:00&07:00&$\SI{-7}{\degree}$&12:00&12:00&$+$04h15m&16:00&16:00&$\SI{02}{\degree}$\\ 
France&FRA&08:00&07:10&$\SI{-6}{\degree}$&13:20&12:30&$+$04h40m&18:00&17:10&$\SI{-9}{\degree}$\\ 
Belgium&BEL&08:00&07:20&$\SI{-7}{\degree}$&12:50&12:10&$+$04h05m&17:00&16:20&$\SI{-3}{\degree}$\\ 
Germany&DEU&07:30&07:05&$\SI{-8}{\degree}$&12:10&11:45&$+$03h45m&16:40&16:15&$\SI{-3}{\degree}$\\ 
Poland&POL&07:00&07:15&$\SI{-7}{\degree}$&12:20&12:35&$+$04h30m&17:00&17:15&$\SI{-12}{\degree}$\\ 
United Kingdom&GBR&08:20&08:15&$\SI{00}{\degree}$&13:00&12:55&$+$04h45m&17:10&17:05&$\SI{-10}{\degree}$\\ 
Lithuania&LIT&07:40&07:15&$\SI{-9}{\degree}$&12:50&12:25&$+$04h00m&18:00&17:35&$\SI{-16}{\degree}$\\ 
Latvia&LVA&08:00&07:35&$\SI{-7}{\degree}$&13:20&12:55&$+$04h15m&18:00&17:35&$\SI{-16}{\degree}$\\ 
Sweden&SWE&07:50&07:55&$\SI{-7}{\degree}$&12:40&12:45&$+$03h45m&16:40&16:45&$\SI{-11}{\degree}$\\ 
Estonia&EST&08:00&07:40&$\SI{-8}{\degree}$&13:10&12:50&$+$03h50m&17:10&16:50&$\SI{-12}{\degree}$\\ 
Norway&NOR&07:50&07:30&$\SI{-10}{\degree}$&12:40&12:20&$+$03h20m&16:00&15:40&$\SI{-5}{\degree}$\\ 
Finland&FIN&07:40&07:20&$\SI{-11}{\degree}$&12:40&12:20&$+$03h05m&16:20&16:00&$\SI{-7}{\degree}$\\ 
\midrule
\multicolumn{2}{l|}{Average}&07:50&07:25&\SI{-6}{\degree}&12:45&12:25&04h10m&17:15&16:50&\SI{-9}{\degree}\\
\multicolumn{2}{l|}{Variability \hfill$2\us(\{x_i\})$}&\SI{45}{\min}&\SI{45}{\min}&\SI{+8}{\degree}&\SI{50}{\min}&\SI{45}{\min}&01h10m&01h45m&01h25m&\SI{+11}{\degree}\\
\bottomrule
\multicolumn{11}{l}{\emph{Both combined}}\\
\midrule
\multicolumn{2}{l|}{Average}&07:55&07:30&\SI{-5}{\degree}&12:50&12:25&04h20m&17:15&16:50&\SI{-8}{\degree}\\
\multicolumn{2}{l|}{Variability \hfill$2\us(\{x_i\})$}&\SI{45}{\min}&\SI{45}{\min}&\SI{8}{\degree}&\SI{45}{\min}&\SI{50}{\min}&01h10m&01h35m&01h15m&\SI{10}{\degree}\\
\bottomrule
\end{tabular}

  \caption{Relevant parameters for labour start, noon and end points; $t$ stands for local time; $\tau$ is mean solar time, $\Delta t_w$ is time distance to winter sunrise  and $z_{w}$ is the winter solar elevation angle at the event. Times have been rounded to the nearest five-minute mark except Irish data which have been rounded to the nearest quarter of an hour. Elevation angles have been rounded to the nearest whole degree. Countries are listed in increasing values of median latitude. Simple descriptive statistics values (sample average value and variability measured as twice sample standard deviation) are listed Data are shown in Figure~\ref{fig:labor}. Correlations are reported on Table~\ref{tab:ajustelabor}.}
  \label{tab:data}
\end{table*}

\begin{table*}
  \centering
\begin{tabular}{l|ccc}
\toprule
\emph{Employees $\phi<\SI{54}{\degree}$}&\textbf{Labor Start}&\textbf{Labor Noon}&\textbf{Labor End}\\
\midrule
\multicolumn{1}{l}{\textbf{Meridional quantity}}&\multicolumn{3}{c}{$y=\tau$}\\
\midrule
vs $x=D_w;$\hfill $r^2$&$\num{0.233}$&$\num{0.277}$&$\num{0.0845}$\\
slope\hfill $p(u_p)$&$\SI{-20.0(97)}{\min\per\hour}$&$\SI{-21.6(93)}{\min\per\hour}$&$\SI{14(13)}{\min\per\hour}$\\
reference\hfill $y_{\text{ref}}$&07:40&12:30&16:50\\
variability\hfill $2\us(\{y_i\})$&$\SI{55}{\min}$&$\SI{55}{\min}$&01h05m\\
observations\hfill $N$&$\num{16}$&$\num{16}$&$\num{16}$\\
\midrule
\multicolumn{1}{l}{\textbf{Latitude prone quantity}}&$y=z_w$&$y=\Delta t_w$ (to sunrise)&$y=z_w$\\
\midrule
vs $x=D_w;$\hfill $r^2$&$\num{0.045}$&$\num{0.0543}$&$\num{0.0211}$\\
slope\hfill $p(u_p)$&$\SI{1.1(14)}{\degree\per\hour}$&$\SI{8.3(93)}{\min\per\hour}$&$\SI{1.0(19)}{\degree\per\hour}$\\
reference\hfill $y_{\text{ref}}$&\ang{-4}&+04h30m&\ang{-8}\\
variability\hfill $2\us(\{y_i\})$&$\SI{7}{\degree}$&$\SI{45}{\min}$&$\SI{10}{\degree}$\\
observations\hfill $N$&$\num{16}$&$\num{16}$&$\num{16}$\\
\toprule
\emph{Employees $\phi>\SI{54}{\degree}$}&&&\\
\midrule
\multicolumn{1}{l}{\textbf{Meridional quantity}}&\multicolumn{3}{c}{$y=\tau$}\\
\midrule
vs $x=D_w;$\hfill $r^2$&$\num{0.00819}$&$\num{0.00326}$&$\num{0.297}$\\
slope\hfill $p(u_p)$&$\SI{-1.8(91)}{\min\per\hour}$&$\SI{ 0(11)}{\min\per\hour}$&$\SI{40(28)}{\min\per\hour}$\\
reference\hfill $y_{\text{ref}}$&07:35&12:35&16:25\\
variability\hfill $2\us(\{y_i\})$&$\SI{25}{\min}$&$\SI{30}{\min}$&01h35m\\
observations\hfill $N$&$\num{7}$&$\num{7}$&$\num{7}$\\
\midrule
\multicolumn{1}{l}{\textbf{Latitude prone quantity}}&$y=z_w$&$y=\Delta t_w$ (to sunrise)&$y=z_w$\\
\midrule
vs $x=D_w;$\hfill $r^2$&$\num{0.363}$&$\num{0.624}$&$\num{0.134}$\\
slope\hfill $p(u_p)$&$\SI{1.63(97)}{\degree\per\hour}$&$\SI{31(11)}{\min\per\hour}$&$\SI{-2.8(33)}{\degree\per\hour}$\\
reference\hfill $y_{\text{ref}}$&\ang{-9}&+03h35m&\ang{-9}\\
variability\hfill $2\us(\{y_i\})$&$\SI{3}{\degree}$&$\SI{50}{\min}$&$\SI{10}{\degree}$\\
observations\hfill $N$&$\num{7}$&$\num{7}$&$\num{7}$\\
\bottomrule\end{tabular}
  \caption{Bivariate correlations for labor time marks and control variables. Top table refers observations below $\ang{54}$ latitude; bottom, observations above that level. On each set the meridional quantity is labor mean solar times ($y=\tau$) and the latitude prone quantity is winter solar elevation angle $z_w$ or distance to winter sunrise $\Delta t_w$ at labor time marks. Either case the quantity is tested against shortest photoperiod $D_w$. Each test reports Pearson's $r^2$ correlation coefficient, slope and its uncertainty $p(u_p)$, a reference value $y_{\text{ref}}$ at the level $D_w=\SI{8}{\hour}$ ($\phi\sim\ang{50}$)  (top) and $D_w=\SI{6}{\hour}$ ($\phi\sim\ang{60}$) (bottom) and the variability of the tested variable computed as twice its sample standard deviation. Uncertainties apply to the least two significant digits. Data are reported on Table~\ref{tab:data} and shown in Figure~\ref{fig:labor}.}
  \label{tab:ajustelabor}
\end{table*}

\begin{table*}
  \centering
\setlength{\tabcolsep}{2pt}
\begin{tabular}{lc|ccr|ccc|cc}
\toprule
\textbf{Country}&\textbf{Label}&\multicolumn{3}{c}{\textbf{Wake up}}&\multicolumn{3}{c}{\textbf{Wakeful noon}}&\multicolumn{2}{c}{\textbf{Bedtime}}\\ 
&&$t$&$\tau$&$z_w$&$t$&$\tau$&$\Delta t_w$&$t$&$\tau$\\ 
\midrule
\multicolumn{10}{l}{\emph{Time Use Surveys (employees in weekday only)}}\\
\midrule
United States&USA&06:00&05:50&$\SI{-16}{\degree}$&14:25&14:15&$+$07h00m&22:25&22:15\\ 
Spain&ESP&07:00&05:50&$\SI{-17}{\degree}$&15:30&14:15&$+$06h55m&23:50&22:35\\ 
Italy&ITA&06:45&06:35&$\SI{-10}{\degree}$&14:55&14:45&$+$07h15m&23:00&22:50\\ 
Canada&CAN&06:20&06:00&$\SI{-16}{\degree}$&14:45&14:30&$+$06h50m&22:40&22:25\\ 
France&FRA&06:40&05:50&$\SI{-19}{\degree}$&14:55&14:05&$+$06h15m&22:55&22:05\\ 
United Kingdom&GBR&06:50&06:45&$\SI{-12}{\degree}$&15:05&14:55&$+$06h45m&23:00&22:55\\ 
Ireland&IRL&07:15&06:45&$\SI{-12}{\degree}$&15:30&15:15&$+$07h00m&23:45&23:15\\ 
Denmark&DNK&06:35&06:20&$\SI{-16}{\degree}$&15:15&15:00&$+$06h30m&23:15&23:05\\ 
\midrule
\multicolumn{2}{l|}{Average}&06:40&06:15&\SI{-15}{\degree}&15:00&14:35&06h50m&23:05&22:40\\
\multicolumn{2}{l|}{Variability \hfill$2\us(\{x_i\})$}&\SI{45}{\min}&\SI{50}{\min}&\SI{+6}{\degree}&\SI{45}{\min}&\SI{45}{\min}&\SI{35}{\min}&\SI{55}{\min}&\SI{45}{\min}\\
\bottomrule
\multicolumn{10}{l}{\emph{Hetus pre-prepared tables (standard population)}}\\
\midrule
Spain&ESP&08:30&07:20&$\SI{-1}{\degree}$&16:10&15:00&$+$07h35m&23:50&22:40\\ 
Bulgaria&BGR&07:20&07:00&$\SI{-6}{\degree}$&15:00&14:40&$+$07h10m&22:30&22:10\\ 
Italy&ITA&07:40&07:30&$\SI{-1}{\degree}$&15:30&15:20&$+$07h45m&23:00&22:50\\ 
Slovenia&SVN&07:00&07:00&$\SI{-7}{\degree}$&14:40&14:40&$+$06h55m&22:00&22:00\\ 
France&FRA&08:00&07:10&$\SI{-6}{\degree}$&15:20&14:30&$+$06h40m&22:50&22:00\\ 
Belgium&BEL&08:00&07:20&$\SI{-7}{\degree}$&15:30&14:50&$+$06h45m&23:00&22:20\\ 
Germany&DEU&07:30&07:05&$\SI{-8}{\degree}$&15:10&14:45&$+$06h45m&22:50&22:25\\ 
Poland&POL&07:20&07:35&$\SI{-5}{\degree}$&14:50&15:05&$+$07h00m&22:10&22:25\\ 
United Kingdom&GBR&07:50&07:45&$\SI{-4}{\degree}$&15:30&15:25&$+$07h15m&23:00&22:55\\ 
Lithuania&LIT&07:10&06:45&$\SI{-13}{\degree}$&14:40&14:15&$+$05h50m&22:00&21:35\\ 
Latvia&LVA&07:30&07:05&$\SI{-11}{\degree}$&15:00&14:35&$+$05h55m&22:30&22:05\\ 
Sweden&SWE&07:30&07:35&$\SI{-9}{\degree}$&15:20&15:25&$+$06h25m&23:00&23:05\\ 
Estonia&EST&07:30&07:10&$\SI{-12}{\degree}$&15:10&14:50&$+$05h50m&22:50&22:30\\ 
Norway&NOR&07:50&07:30&$\SI{-10}{\degree}$&15:40&15:20&$+$06h20m&23:20&23:00\\ 
Finland&FIN&07:30&07:10&$\SI{-13}{\degree}$&15:10&14:50&$+$05h35m&22:50&22:30\\ 
\midrule
\multicolumn{2}{l|}{Average}&07:35&07:15&\SI{-7}{\degree}&15:15&14:55&06h40m&22:45&22:25\\
\multicolumn{2}{l|}{Variability \hfill$2\us(\{x_i\})$}&\SI{45}{\min}&\SI{35}{\min}&\SI{+7}{\degree}&\SI{50}{\min}&\SI{45}{\min}&01h20m&01h00m&\SI{50}{\min}\\
\bottomrule
\multicolumn{10}{l}{\emph{Both combined}}\\
\midrule
\multicolumn{2}{l|}{Average}&&&&&&&22:55&22:30\\
\multicolumn{2}{l|}{Variability \hfill$2\us(\{x_i\})$}&&&&&&&01h00m&\SI{50}{\min}\\
\bottomrule
\end{tabular}
  \caption{Relevant parameters for wakeup time, wakeful noon time and bedtime; $t$ stands for local time; $\tau$ is mean solar time, $\Delta t_w$ is time distance to winter sunrise and $z_{w}$ is the winter solar elevation angle at the event. Hetus data were obtained from the ``sleep and other personal care'' daily rhythm. Times have been rounded to the nearest five-minute mark except Irish data which have been rounded to the nearest quarter of an hour. Elevation angles have been rounded to the nearest whole degree. Simple descriptive statistic values (sample average value and twice sample starndard deviation) are listed. Data are shown in Figure~\ref{fig:vigil}. Bivariate correlations are reported on Table~\ref{tab:ajusteVigil}.}
  \label{tab:vigil}
\end{table*}

\begin{table*}
  \centering
\begin{tabular}{l|ccc}\toprule
\emph{Employees}&\textbf{Wake up}&\textbf{Wakeful noon}&\textbf{Bedtime}\\
\midrule
\multicolumn{1}{l}{\textbf{Meridional quantity}}&\multicolumn{3}{c}{$y=\tau$}\\
\midrule
vs $x=D_w;$\hfill $r^2$&$\num{0.407}$&$\num{0.6}$&$\num{0.443}$\\
slope\hfill $p(u_p)$&$\SI{-17.8(88)}{\min\per\hour}$&$\SI{-20.3(68)}{\min\per\hour}$&$\SI{-17.1(79)}{\min\per\hour}$\\
reference\hfill $y_{\text{ref}}$&06:20&14:45&22:50\\
variability\hfill $2\us(\{y_i\})$&$\SI{50}{\min}$&$\SI{45}{\min}$&$\SI{45}{\min}$\\
observations\hfill $N$&$\num{8}$&$\num{8}$&$\num{8}$\\
\midrule
\multicolumn{1}{l}{\textbf{Latitude prone quantity}}&$y=z_w$&$y=\Delta t_w$ (to sunrise)&$y=\Delta t_w$ (to sunrise)\\
\midrule
vs $x=D_w;$\hfill $r^2$&$\num{0.0292}$&$\num{0.253}$&$\num{0.307}$\\
slope\hfill $p(u_p)$&$\SI{0.0(14)}{\degree\per\hour}$&$\SI{9.6(68)}{\min\per\hour}$&$\SI{12.8(79)}{\min\per\hour}$\\
reference\hfill $y_{\text{ref}}$&\ang{-15}&+06h45m&$-$09h10m\\
variability\hfill $2\us(\{y_i\})$&$\SI{6}{\degree}$&$\SI{35}{\min}$&$\SI{40}{\min}$\\
observations\hfill $N$&$\num{8}$&$\num{8}$&$\num{8}$\\
\toprule
\emph{Starndard population}&&&\\
\midrule
\multicolumn{1}{l}{\textbf{Meridional quantity}}&\multicolumn{3}{c}{$y=\tau$}\\
\midrule
vs $x=D_w;$\hfill $r^2$&$\num{0.0131}$&$\num{0.0276}$&$\num{0.07}$\\
slope\hfill $p(u_p)$&$\SI{-1.5(37)}{\min\per\hour}$&$\SI{-2.8(47)}{\min\per\hour}$&$\SI{-5.3(54)}{\min\per\hour}$\\
reference\hfill $y_{\text{ref}}$&07:15&14:50&22:25\\
variability\hfill $2\us(\{y_i\})$&$\SI{35}{\min}$&$\SI{45}{\min}$&$\SI{50}{\min}$\\
observations\hfill $N$&$\num{15}$&$\num{15}$&$\num{15}$\\
\midrule
\multicolumn{1}{l}{\textbf{Latitude prone quantity}}&$y=z_w$&$y=\Delta t_w$ (to sunrise)&$y=\Delta t_w$ (to sunrise)\\
\midrule
vs $x=D_w;$\hfill $r^2$&$\num{0.634}$&$\num{0.721}$&$\num{0.612}$\\
slope\hfill $p(u_p)$&$\SI{2.36(50)}{\degree\per\hour}$&$\SI{27.1(47)}{\min\per\hour}$&$\SI{24.6(54)}{\min\per\hour}$\\
reference\hfill $y_{\text{ref}}$&\ang{-6}&+06h50m&$-$09h35m\\
variability\hfill $2\us(\{y_i\})$&$\SI{7}{\degree}$&01h20m&01h20m\\
observations\hfill $N$&$\num{15}$&$\num{15}$&$\num{15}$\\
\toprule
\emph{Both combined}&\textbf{\phantom{Wake up}}&\textbf{\phantom{Wakeful noon}}&\textbf{\phantom{Bedtime}}\\
\midrule
\multicolumn{1}{l}{\textbf{Meridional quantity}}&\multicolumn{3}{c}{$y=\tau$}\\
\midrule
vs $x=D_w;$\hfill $r^2$&&&$\num{0.0466}$\\
slope\hfill $p(u_p)$&&&$\SI{-4.5(45)}{\min\per\hour}$\\
reference\hfill $y_{\text{ref}}$&&&22:30\\
variability\hfill $2\us(\{y_i\})$&&&$\SI{50}{\min}$\\
observations\hfill $N$&&&$\num{23}$\\
\midrule
\multicolumn{1}{l}{\textbf{Latitude prone quantity}}&&&$y=\Delta t_w$ (to sunrise)\\
\midrule
vs $x=D_w;$\hfill $r^2$&&&$\num{0.599}$\\
slope\hfill $p(u_p)$&&&$\SI{25.4(45)}{\min\per\hour}$\\
reference\hfill $y_{\text{ref}}$&&&$-$09h30m\\
variability\hfill $2\us(\{y_i\})$&&&01h20m\\
observations\hfill $N$&&&$\num{23}$\\
\bottomrule
\end{tabular}
  \caption{Bivariate correlations for sleep-wake time marks and control variables and subsets. Top table reports data for employees; next standard population (Hetus); finally bedtimes of both sets combined. The meridional quantity is mean solar time at time marks ($y=\tau$) and the latitude prone quantity is winter solar elevation angle $z_w$ or distance to winter sunrise $\Delta t_w$ at sleep-wake time marks. Either case the quantity is tested against shortest photoperiod $D_w$. Each test reports Pearson's $r^2$ correlation coefficient, slope and its uncertainty $p(u_p)$, a reference value $y_{\text{ref}}$ at the level $D_w=\SI{8}{\hour}$ ($\phi\sim\ang{50}$) and the variability of the tested variable computed as twice its sample standard deviation. Uncertainties apply to the least two significant digits.}
  \label{tab:ajusteVigil}
\end{table*}

\begin{table*}
  \centering
\setlength{\tabcolsep}{2pt}
\begin{tabular}{lc|cc}
\toprule
\textbf{Country}&\textbf{Label}&\multicolumn{2}{c}{\textbf{TV Prime Time}}\\ 
&&$t$&$\tau$\\ 
\midrule
\multicolumn{4}{l}{\emph{Time Use Surveys (employees in weekday only)}}\\
\midrule
United States&USA&20:45&20:40\\ 
Spain&ESP&22:45&21:35\\ 
Italy&ITA&21:45&21:35\\ 
Canada&CAN&21:30&21:15\\ 
France&FRA&21:45&20:55\\ 
United Kingdom&GBR&21:30&21:25\\ 
Ireland&IRL&22:30&22:00\\ 
Denmark&DNK&21:30&21:15\\ 
\midrule
\multicolumn{2}{l|}{Average}&21:45&21:20\\
\multicolumn{2}{l|}{Variability \hfill$2\us(\{x_i\})$}&01h15m&\SI{50}{\min}\\
\bottomrule
\multicolumn{4}{l}{\emph{Hetus pre-prepared tables (standard population)}}\\
\midrule
Spain&ESP&22:40&21:30\\ 
Bulgaria&BGR&21:30&21:10\\ 
Italy&ITA&21:40&21:30\\ 
Slovenia&SVN&20:50&20:50\\ 
France&FRA&21:50&21:00\\ 
Belgium&BEL&21:50&21:10\\ 
Germany&DEU&21:20&20:55\\ 
Poland&POL&20:50&21:05\\ 
United Kingdom&GBR&21:40&21:35\\ 
Lithuania&LIT&20:40&20:15\\ 
Latvia&LVA&20:50&20:25\\ 
Sweden&SWE&21:20&21:25\\ 
Estonia&EST&21:10&20:50\\ 
Norway&NOR&21:20&21:00\\ 
Finland&FIN&21:40&21:20\\ 
\midrule
\multicolumn{2}{l|}{Average}&21:25&21:05\\
\multicolumn{2}{l|}{Variability \hfill$2\us(\{x_i\})$}&01h00m&\SI{45}{\min}\\
\bottomrule
\multicolumn{4}{l}{\emph{Both combined}}\\
\midrule
\multicolumn{2}{l|}{Average}&21:30&21:10\\
\multicolumn{2}{l|}{Variability \hfill$2\us(\{x_i\})$}&01h10m&\SI{50}{\min}\\
\bottomrule
\end{tabular}
  \caption{Prime time mark obtained from the peak position of the daily watching TV daily rhythm. The table lists local time ($t$) and mean solar time ($\tau$). Simple descriptive statistic values (sample average value and twice sample standard deviation) are listed.  Times have been rounded to the nearest fit-minute mark except Irish data which have been rounded to the next quarter of an hour. Bivariate correlations are reported on Table~\ref{tab:ociodata}.}
  \label{tab:ociodata}
\end{table*}

\begin{table*}
  \centering
\begin{tabular}{l|c}\toprule
\emph{All combined}&\textbf{Prime Time}\\
\midrule
\multicolumn{1}{l}{\textbf{Meridional quantity}}&\multicolumn{1}{c}{$y=\tau$}\\
\midrule
vs $x=D_w;$\hfill $r^2$&$\num{0.0342}$\\
slope\hfill $p(u_p)$&$\SI{3.8(45)}{\min\per\hour}$\\
reference\hfill $y_{\text{ref}}$&21:10\\
variability\hfill $2\us(\{y_i\})$&$\SI{50}{\min}$\\
observations\hfill $N$&$\num{23}$\\
\midrule
\multicolumn{1}{l}{\textbf{Latitude prone quantity}}&$y=\Delta t_w$ (to sunset)\\
\midrule
vs $x=D_w;$\hfill $r^2$&$\num{0.619}$\\
slope\hfill $p(u_p)$&$\SI{-26.1(45)}{\min\per\hour}$\\
reference\hfill $y_{\text{ref}}$&+05h10m\\
variability\hfill $2\us(\{y_i\})$&01h20m\\
observations\hfill $N$&$\num{23}$\\
\bottomrule\end{tabular}
  \caption{Bivariate correlations for  TV prime time marks and control variables for the full set of data. The meridional quantity is labor mean solar times ($y=\tau$) the latitude prone quantity is distance to winter sunrise $\Delta t_w$ at TV prime time mark. Either case the quantity is tested against shortest photoperiod $D_w$. Each test reports Pearson's $r^2$ correlation coefficient, slope and its uncertainty $p(u_p)$, a reference value $y_{\text{ref}}$ at the level $D_w=\SI{8}{\hour}$ ($\phi\sim\ang{50}$) and the variability of the tested variable measured as twice its sample standard deviation. Last two items were rounded to the nearest fifth minute. Uncertainties apply to the least two significant digits. Data are listed in Table~\ref{tab:ociodata}.}
  \label{tab:ajusteTV}
\end{table*}

\begin{table*}
  \centering
\setlength{\tabcolsep}{2pt}
\begin{tabular}{lc|ccc|ccc}
\toprule
\textbf{Country}&\textbf{Label}&\multicolumn{3}{c}{\textbf{Leaving Home}}&\multicolumn{3}{c}{\textbf{Coming Home}}\\ 
&&$t$&$\tau$&$z_w$&$t$&$\tau$&$\Delta t_w$\\ 
\midrule
\multicolumn{8}{l}{\emph{Time Use Surveys (employees in weekday only)}}\\
\midrule
United States&USA&07:00&06:50&$\SI{-5}{\degree}$&18:50&18:40&$+$01h55m\\ 
Spain&ESP&07:45&06:30&$\SI{-9}{\degree}$&20:30&19:20&$+$02h40m\\ 
Italy&ITA&07:30&07:20&$\SI{-3}{\degree}$&19:15&19:05&$+$02h35m\\ 
Canada&CAN&07:20&07:00&$\SI{-7}{\degree}$&18:20&18:00&$+$01h40m\\ 
France&FRA&07:35&06:45&$\SI{-10}{\degree}$&18:55&18:05&$+$01h55m\\ 
United Kingdom&GBR&07:45&07:35&$\SI{-5}{\degree}$&18:15&18:05&$+$02h15m\\ 
Denmark&DNK&07:25&07:10&$\SI{-10}{\degree}$&17:50&17:35&$+$02h05m\\ 
\midrule
\multicolumn{2}{l|}{Average}&07:30&07:05&\SI{-7}{\degree}&18:45&18:20&02h10m\\
\multicolumn{2}{l|}{Variability \hfill$2\us(\{x_i\})$}&\SI{40}{\min}&\SI{50}{\min}&\SI{+5}{\degree}&01h40m&01h15m&\SI{40}{\min}\\
\bottomrule
\end{tabular}
\caption{Leaving home/coming home time marks obtained from the shares of employees not located at home as a function of time; $t$ stands for local time; $\tau$ is mean solar time, $\Delta t_w$ is time distance to winter sunset;  and $z_{w}$ is the winter solar elevation angle at the event. Times have been rounded to the nearest fifth-minute mark except Irish data which have been rounded to the next quarter of an hour. Simple descriptive statistic values (sample average value and twice sample standard deviation) are listed. Bivariate correlations are reported on Table~\ref{tab:ajusteCasa}.}
  \label{tab:casa}
\end{table*}

\begin{table*}
  \centering
\begin{tabular}{l|cc}\toprule
\emph{Employees}&\textbf{Leaving Home}&\textbf{Coming Home}\\
\midrule
\multicolumn{1}{l}{\textbf{Meridional quantity}}&\multicolumn{2}{c}{$y=\tau$}\\
\midrule
vs $x=D_w;$\hfill $r^2$&$\num{0.369}$&$\num{0.635}$\\
slope\hfill $p(u_p)$&$\SI{-15.5(91)}{\min\per\hour}$&$\SI{34(12)}{\min\per\hour}$\\
reference\hfill $y_{\text{ref}}$&07:10&18:10\\
variability\hfill $2\us(\{y_i\})$&$\SI{45}{\min}$&01h15m\\
observations\hfill $N$&$\num{7}$&$\num{7}$\\
\midrule
\multicolumn{1}{l}{\textbf{Latitude prone quantity}}&$y=z_w$&$y=\Delta t_w$ (to sunset)\\
\midrule
vs $x=D_w;$\hfill $r^2$&$\num{0.0705}$&$\num{0.028}$\\
slope\hfill $p(u_p)$&$\SI{0.0(14)}{\degree\per\hour}$&$\SI{ 0(12)}{\min\per\hour}$\\
reference\hfill $y_{\text{ref}}$&\ang{-7}&+02h10m\\
variability\hfill $2\us(\{y_i\})$&$\SI{6}{\degree}$&$\SI{45}{\min}$\\
observations\hfill $N$&$\num{7}$&$\num{7}$\\
\bottomrule\end{tabular}
  \caption{Bivariate correlations for leaving home/coming home time marks and control variables. The meridional quantity is mean solar times ($y=\tau$) at leaving home/coming home marks, and the latitude prone quantity is winter solar elevation angle $z_w$ or distance to winter sunset $\Delta t_w$ at these time marks. Either case the quantity is tested against shortest photoperiod $D_w$. Each test reports Pearson's $r^2$ correlation coefficient, slope and its uncertainty $p(u_p)$, a reference value $y_{\text{ref}}$ at the level $D_w=\SI{8}{\hour}$ ($\phi\sim\ang{50}$ and the variability of the tested variable measured as twice its sample standard population. Last two items were rounded to the nearest fifth minute. Uncertainties apply to the least two significant digits. Data are listed in Table~\ref{tab:casa}.}
  \label{tab:ajusteCasa}
\end{table*}

\begin{table*}
  \centering
\setlength{\tabcolsep}{2pt}
\begin{tabular}{lc|ccr|ccc|ccc}
\toprule
\textbf{Country}&\textbf{Label}&\multicolumn{3}{c}{\textbf{Breakfast}}&\multicolumn{3}{c}{\textbf{Lunch}}&\multicolumn{3}{c}{\textbf{Dinner}}\\ 
&&$t$&$\tau$&$z_w$&$t$&$\tau$&$\Delta t_w$&$t$&$\tau$&$\Delta t_w$\\ 
\midrule
\multicolumn{11}{l}{\emph{Time Use Surveys (employees in weekday only)}}\\
\midrule
United States&USA&07:00&06:55&$\SI{-5}{\degree}$&12:15&12:05&$-$04h35m&18:40&18:30&$+$01h45m\\ 
Spain&ESP&07:30&06:15&$\SI{-12}{\degree}$&14:25&13:15&$-$03h25m&21:25&20:15&$+$03h35m\\ 
Italy&ITA&07:20&07:10&$\SI{-4}{\degree}$&13:15&13:05&$-$03h20m&20:15&20:05&$+$03h35m\\ 
Canada&CAN&07:05&06:50&$\SI{-8}{\degree}$&12:20&12:00&$-$04h20m&18:10&17:55&$+$01h35m\\ 
France&FRA&07:15&06:25&$\SI{-13}{\degree}$&12:40&11:50&$-$04h25m&20:05&19:15&$+$03h05m\\ 
United Kingdom&GBR&07:30&07:25&$\SI{-7}{\degree}$&13:00&12:50&$-$03h00m&18:40&18:35&$+$02h45m\\ 
Ireland&IRL&08:15&07:45&$\SI{-4}{\degree}$&13:15&12:45&$-$03h00m&18:30&18:00&$+$02h15m\\ 
Denmark&DNK&07:20&07:05&$\SI{-10}{\degree}$&12:20&12:05&$-$03h25m&18:30&18:15&$+$02h45m\\ 
\midrule
\multicolumn{2}{l|}{Average}&07:25&07:00&\SI{-8}{\degree}&12:55&12:30&03h40m&19:15&18:50&02h40m\\
\multicolumn{2}{l|}{Variability \hfill$2\us(\{x_i\})$}&\SI{45}{\min}&01h00m&\SI{+7}{\degree}&01h30m&01h05m&01h20m&02h20m&01h45m&01h30m\\
\bottomrule
\multicolumn{11}{l}{\emph{Hetus pre-prepared tables (standard population)}}\\
\midrule
Spain&ESP&09:20&08:10&$\SI{06}{\degree}$&14:30&13:20&$-$03h20m&21:30&20:20&$+$03h40m\\ 
Bulgaria&BGR&08:30&08:10&$\SI{05}{\degree}$&12:30&12:10&$-$04h20m&19:40&19:20&$+$02h50m\\ 
Italy&ITA&07:50&07:40&$\SI{00}{\degree}$&13:10&13:00&$-$03h30m&20:20&20:10&$+$03h40m\\ 
Slovenia&SVN&08:20&08:20&$\SI{04}{\degree}$&12:50&12:50&$-$03h30m&19:10&19:10&$+$02h50m\\ 
France&FRA&07:50&07:00&$\SI{-8}{\degree}$&12:40&11:50&$-$04h20m&19:50&19:00&$+$02h50m\\ 
Belgium&BEL&08:30&07:50&$\SI{-3}{\degree}$&12:30&11:50&$-$04h10m&18:50&18:10&$+$02h10m\\ 
Germany&DEU&08:40&08:15&$\SI{01}{\degree}$&12:30&12:05&$-$03h50m&18:50&18:25&$+$02h30m\\ 
Poland&POL&08:30&08:45&$\SI{04}{\degree}$&14:00&14:15&$-$01h35m&19:10&19:25&$+$03h35m\\ 
United Kingdom&GBR&08:10&08:05&$\SI{-2}{\degree}$&13:00&12:55&$-$02h55m&18:10&18:05&$+$02h15m\\ 
Lithuania&LIT&08:30&08:05&$\SI{-3}{\degree}$&13:00&12:35&$-$03h00m&19:10&18:45&$+$03h10m\\ 
Latvia&LVA&07:30&07:05&$\SI{-11}{\degree}$&12:50&12:25&$-$02h55m&19:10&18:45&$+$03h25m\\ 
Sweden&SWE&09:00&09:05&$\SI{00}{\degree}$&12:20&12:25&$-$02h40m&18:20&18:25&$+$03h20m\\ 
Estonia&EST&08:20&08:00&$\SI{-6}{\degree}$&13:10&12:50&$-$02h10m&18:50&18:30&$+$03h30m\\ 
Norway&NOR&09:10&08:50&$\SI{-2}{\degree}$&11:40&11:20&$-$03h35m&16:40&16:20&$+$01h25m\\ 
Finland&FIN&08:20&08:00&$\SI{-7}{\degree}$&11:40&11:20&$-$03h25m&17:10&16:50&$+$02h05m\\ 
\midrule
\multicolumn{2}{l|}{Average}&08:25&08:05&\SI{-1}{\degree}&12:50&12:30&03h15m&19:00&18:40&02h55m\\
\multicolumn{2}{l|}{Variability \hfill$2\us(\{x_i\})$}&01h00m&01h10m&\SI{+10}{\degree}&01h30m&01h30m&01h35m&02h20m&02h05m&01h20m\\
\bottomrule
\multicolumn{11}{l}{\emph{Both combined}}\\
\midrule
\multicolumn{2}{l|}{Average}&08:05&07:40&\SI{-4}{\degree}&12:50&12:30&03h25m&19:05&18:45&02h50m\\
\multicolumn{2}{l|}{Variability \hfill$2\us(\{x_i\})$}&01h20m&01h30m&\SI{11}{\degree}&01h25m&01h25m&01h30m&02h20m&02h00m&01h25m\\
\bottomrule
\end{tabular}\caption{Relevant pararameters for breakfast, lunch and dinner. For each subset the table lists local time ($t$), mean solar time ($\tau$), distance ($\Delta t_w$) to winter sunset (lunch and dinner) and winter solar elevation angle $z_w$. Times have been rounded to the nearest five-minute mark except Irish data which have been rounded to the nearest quarter of an hour, angles have been rounded to whole numbers. Data are shown in Figure~\ref{fig:meal}. Bivariate correlations are reported on Table~\ref{tab:ajusteComida}.}
  \label{tab:meal}
\end{table*}

\begin{table*}
  \centering
\begin{tabular}{l|cc}\toprule
\emph{Europe}&\textbf{Lunch}&\textbf{Dinner}\\
\midrule
\multicolumn{1}{l}{\textbf{Meridional quantity}}&\multicolumn{2}{c}{$y=\tau$}\\
\midrule
vs $x=D_w;$\hfill $r^2$&$\num{0.187}$&$\num{0.659}$\\
slope\hfill $p(u_p)$&$\SI{15.5(77)}{\min\per\hour}$&$\SI{38.4(65)}{\min\per\hour}$\\
reference\hfill $y_{\text{ref}}$&12:40&19:00\\
variability\hfill $2\us(\{y_i\})$&01h25m&01h45m\\
observations\hfill $N$&$\num{20}$&$\num{20}$\\
\midrule
\multicolumn{1}{l}{\textbf{Latitude prone quantity}}&$y=\Delta t_w$ (to sunset)&$y=\Delta t_w$ (to sunset)\\
\midrule
vs $x=D_w;$\hfill $r^2$&$\num{0.165}$&$\num{0.0859}$\\
slope\hfill $p(u_p)$&$\SI{-14.4(77)}{\min\per\hour}$&$\SI{8.4(65)}{\min\per\hour}$\\
reference\hfill $y_{\text{ref}}$&-03h20m&+03h00m\\
variability\hfill $2\us(\{y_i\})$&01h25m&01h05m\\
observations\hfill $N$&$\num{20}$&$\num{20}$\\
\bottomrule\end{tabular}
  \caption{Bivariate correlations for lunch and dinner time marks and control variables for European countries. The meridional quantity is mean solar times ($y=\tau$) at main meals and the latitude prone quantity is distance to winter sunset $\Delta t_w$ at main meals. Either case the quantity is tested against shortest photoperiod $D_w$. Each test reports Pearson's $r^2$ correlation coefficient, slope and its uncertainty $p(u_p)$, the predicted value $y_{\text{ref}}$ at the level $D_w=\SI{8}{\hour}$ ($\phi\sim\ang{50}$) and the variability of the tested variable measured as twice its sample standard population. Last two items were rounded to the nearest fifth minute. Uncertainties apply to the least two significant digits. Data are listed in Table~\ref{tab:meal} and shown in Figure~\ref{fig:meal}. Polish lunch time and Norwegian dinner time did not enter in the correlation analysis.}
  \label{tab:ajusteComida}
\end{table*}

\begin{table*}
  \centering
\setlength{\tabcolsep}{2pt}
\begin{tabular}{lc|cc|c|c|c}
\toprule
\textbf{Country}&\textbf{Label}&\multicolumn{2}{c}{Labor duration}&Sleep time&Meal time&TV time\\ 
&&$L$&$\Delta L$&$S$&$M$&$TV$\\ 
\midrule
\multicolumn{5}{l}{\emph{Time Use Survey (employees in weekday only)}}\\
\midrule
United States&USA&07h55m&$-$01h35m&07h45m&01h00m&02h40m\\ 
Spain&ESP&07h45m&$-$01h35m&07h30m&01h40m&01h30m\\ 
Italy&ITA&07h40m&$-$01h15m&07h40m&01h45m&01h30m\\ 
Canada&CAN&07h55m&$-$00h45m&07h45m&01h05m&01h45m\\ 
France&FRA&06h55m&$-$01h25m&07h35m&01h55m&02h20m\\ 
United Kingdom&GBR&07h25m&$-$00h15m&07h45m&01h15m&01h55m\\ 
Ireland&IRL&06h45m&$-$00h45m&07h30m&01h45m&01h45m\\ 
Denmark&DNK&06h25m&$-$00h35m&07h25m&01h50m&01h55m\\ 
\midrule
\multicolumn{2}{l|}{Average}&07h20m&01h00m&07h35m&01h30m&01h55m\\
\multicolumn{2}{l|}{Variability \hfill$2\us$}&01h10m&01h00m&$\SI{20}{\min}$&$\SI{40}{\min}$&$\SI{50}{\min}$\\
\bottomrule
\multicolumn{5}{l}{\emph{Hetus pre-prepared tables (full population set)}}\\
\midrule
Spain&ESP&07h50m&$-$01h25m&08h35m&01h45m&02h50m\\ 
Bulgaria&BGR&08h10m&$-$00h50m&09h05m&02h00m&02h50m\\ 
Italy&ITA&07h35m&$-$01h20m&08h20m&01h55m&02h00m\\ 
Slovenia&SVN&07h30m&$-$01h10m&08h20m&01h30m&02h25m\\ 
France&FRA&07h10m&$-$01h10m&08h50m&02h15m&03h10m\\ 
Belgium&BEL&07h20m&$-$00h35m&08h25m&01h50m&02h55m\\ 
Germany&DEU&07h15m&$-$00h40m&08h10m&01h45m&02h30m\\ 
Poland&POL&07h15m&$-$00h35m&08h30m&01h35m&02h25m\\ 
United Kingdom&GBR&07h25m&$-$00h15m&08h25m&01h25m&02h55m\\ 
Lithuania&LIT&07h50m&$+$00h40m&08h30m&01h30m&02h45m\\ 
Latvia&LVA&08h10m&$+$01h30m&08h40m&01h30m&02h40m\\ 
Sweden&SWE&07h40m&$+$01h35m&08h05m&01h35m&02h10m\\ 
Estonia&EST&08h05m&$+$02h00m&08h25m&01h15m&02h25m\\ 
Norway&NOR&07h10m&$+$01h15m&08h05m&01h20m&02h35m\\ 
Finland&FIN&07h30m&$+$02h05m&08h25m&01h20m&03h00m\\ 
\midrule
\multicolumn{2}{l|}{Average}&07h35m&$\SI{05}{\min}$&08h25m&01h40m&02h40m\\
\multicolumn{2}{l|}{Variability \hfill$2\us$}&$\SI{45}{\min}$&02h35m&$\SI{35}{\min}$&$\SI{35}{\min}$&$\SI{40}{\min}$\\
\bottomrule
\end{tabular}
\caption{Daily average times of primary activities laboring $L$, sleeping $S$ and eating $M$, and TV watching $T$. Also $\Delta L$ lists the time difference of the labor time and the shortest photoperiod $D_w$. Hetus data were retrieved from employment daily totals in main activities pre-prepared table taking into account its participation rate, and sleep daily totals in main activities 2-digit level pre-prepared tables. Times have been rounded to the nearest five-minute mark except Irish data which have been rounded to the next quarter of an hour. For each set average value and variability defined as twice sample standard deviation are reported. Data are shown in Figure~\ref{fig:laborConsump}, bivariate correlations are shown in Table~\ref{tab:ajusteDuracion}.}
  \label{tab:duracion}
\end{table*}

\begin{table*}
  \centering
\begin{tabular}{l|cccc}\toprule
&\textbf{Labor time}&\textbf{Sleep time}&\textbf{Eat time}&\textbf{TV time}\\
\midrule
\multicolumn{5}{l}{\emph{Time Use Surveys (employees in weekday only)}}\\
\midrule
vs $x=D_w;$\hfill $r^2$&$\num{0.743}$&$\num{0.242}$&$\num{0.121}$&$\num{0.0202}$\\
slope\hfill $p(u_p)$&$\SI{33.5(81)}{\min\per\hour}$&$\SI{4.8(35)}{\min\per\hour}$&$\SI{-8.0(88)}{\min\per\hour}$&$\SI{ 0(11)}{\min\per\hour}$\\
reference\hfill $y_{\text{ref}}$&07h10m&07h35m&01h35m&01h55m\\
variability\phantom{adfja}\hfill $2\us(\{y_i\})$&01h10m&$\SI{20}{\min}$&$\SI{40}{\min}$&$\SI{50}{\min}$\\
observations\hfill $N$&$\num{8}$&$\num{8}$&$\num{8}$&$\num{8}$\\
\midrule
\multicolumn{5}{l}{\emph{Hetus pre-prepared tables (full population set)}}\\
\midrule
vs $x=D_w;$\hfill $r^2$&$\num{3.68e-06}$&$\num{0.197}$&$\num{0.513}$&$\num{0.000716}$\\
slope\hfill $p(u_p)$&$\SI{0.0(48)}{\min\per\hour}$&$\SI{5.8(33)}{\min\per\hour}$&$\SI{9.5(26)}{\min\per\hour}$&$\SI{0.0(42)}{\min\per\hour}$\\
reference\hfill $y_{\text{ref}}$&07h35m&08h30m&01h40m&02h40m\\
variability\phantom{adfja}\hfill $2\us(\{y_i\})$&$\SI{45}{\min}$&$\SI{35}{\min}$&$\SI{35}{\min}$&$\SI{40}{\min}$\\
observations\hfill $N$&$\num{15}$&$\num{15}$&$\num{15}$&$\num{15}$\\
\midrule
\multicolumn{5}{l}{\emph{Both combined}}\\
\midrule
vs $x=D_w;$\hfill $r^2$&$\num{0.0274}$&&&\\
slope\hfill $p(u_p)$&$\SI{3.7(49)}{\min\per\hour}$&&&\\
reference\hfill $y_{\text{ref}}$&07h30m&&&\\
variability\phantom{adfja}\hfill $2\us(\{y_i\})$&$\SI{55}{\min}$&&&\\
observations\hfill $N$&$\num{23}$&&&\\
\midrule
\multicolumn{5}{l}{\emph{Eurostat Weekly hours}}\\
\midrule
vs $x=D_w;$\hfill $r^2$&$\num{0.232}$&&&\\
slope\hfill $p(u_p)$&$\SI{15.2(51)}{\min\per\hour}$&&&\\
reference\hfill $y_{\text{ref}}$&07h35m&&&\\
variability\phantom{adfja}\hfill $2\us(\{y_i\})$&01h15m&&&\\
observations\hfill $N$&$\num{32}$&&&\\
\midrule
\multicolumn{5}{l}{\emph{All combined}}\\
\midrule
vs $x=D_w;$\hfill $r^2$&$\num{0.14}$&&&\\
slope\hfill $p(u_p)$&$\SI{10.5(36)}{\min\per\hour}$&&&\\
reference\hfill $y_{\text{ref}}$&07h35m&&&\\
variability\phantom{adfja}\hfill $2\us(\{y_i\})$&01h05m&&&\\
observations\hfill $N$&$\num{55}$&&&\\
\midrule
\multicolumn{5}{l}{\emph{All combined and $\phi<\ang{54}$}}\\
\midrule
vs $x=D_w;$\hfill $r^2$&$\num{0.338}$&&&\\
slope\hfill $p(u_p)$&$\SI{28.8(65)}{\min\per\hour}$&&&\\
reference\hfill $y_{\text{ref}}$&07h20m&&&\\
variability\phantom{adfja}\hfill $2\us(\{y_i\})$&01h05m&&&\\
observations\hfill $N$&$\num{41}$&&&\\
\bottomrule
\end{tabular}

  \caption{Bivariate correlations for average daily consumptions versus shortest photoperiod. Each test reports Pearson's $r^2$ correlation coefficient, slope and its uncertainty $p(u_p)$, the predicted value $y_{\text{ref}}$ at the level $D_w=\SI{8}{\hour}$ ($\phi\sim\ang{50}$) and the variability of the tested variable measured as twice its sample standard population. Last two items were rounded to the nearest fifth minute. Uncertainties apply to the least two significant digits. Different sets and combinations are tested for labor time. Table~\ref{tab:duracion} lists values and Figure~\ref{fig:laborConsump} shows labor, sleep and eat times.}
  \label{tab:ajusteDuracion}
\end{table*}

\end{document}